\documentclass[11pt]{article}

\usepackage{amsmath,amsfonts,amssymb,amsthm,bbm,stmaryrd,mathtools}
\usepackage[dvips]{graphicx}
\usepackage{latexsym}
\usepackage{psfrag}
\usepackage[latin1]{inputenc}
\usepackage[hidelinks]{hyperref}
\usepackage{color}
\numberwithin{equation}{section}
\usepackage{float}
\usepackage{multirow}
\usepackage{upgreek}

\newcommand{\be}{\begin{equation}}
\newcommand{\ee}{\end{equation}}
\newcommand{\barray}{\begin{array}}
\newcommand{\earray}{\end{array}}
\newcommand{\bea}{\begin{eqnarray}}
\newcommand{\eea}{\end{eqnarray}}
\newcommand{\bs}{\begin{subequations}}
\newcommand{\es}{\end{subequations}}

\newcommand{\bit}{\begin{itemize}}
\newcommand{\eit}{\end{itemize}}
\newcommand{\bd}{\begin{description}}
\newcommand{\ed}{\end{description}}

\def\nn{\nonumber}

\def\w{\wedge}
\def\cw{\curlywedge}
\newcommand{\p}{\partial}

\newcommand{\R}{\mathbb{R}}

\newcommand{\f}{\frac}

  \newcommand{\g}{\gamma}  
\renewcommand{\d}{\delta}  \newcommand{\eps}{\epsilon} 
 \renewcommand{\th}{\theta}  \newcommand{\vth}{\vartheta} 
  \renewcommand{\k}{\kappa}  \renewcommand{\l}{\lambda}
\let\m=\mu    \let\n=\nu    \let\om=\omega
 \newcommand{\s}{\sigma}      
     
 \let\Om=\Omega

\def\cH{{\cal H}}

\def\cN{{\cal N}}

\newcommand{\norm}[1]{|\!|#1|\!|}

\newcommand{\eqons}{\,\hat{=}\,}

\usepackage{pgf}
\makeatletter
\newcommand*{\pgfunderleftarrow}{%
  \@ifstar
    {\let\ifpgf@depth\iftrue\mathpalette\@pgfunderleftarrow}
    {\let\ifpgf@depth\iffalse\mathpalette\@pgfunderleftarrow}%
}
\newcommand*{\@pgfunderleftarrow}[2]{%
  #2%
  \edef\pgf@math@fam{\the\fam}%
  \pgfpicture
    \pgfsetbaseline{0pt}
    \pgf@relevantforpicturesizefalse      
    \pgfsetroundcap                       
    \pgfsetarrowsend{to}
    \pgfutil@tempdima=0.28pt%
    \advance\pgfutil@tempdima by.8\pgflinewidth%
    \pgfutil@tempdima-4\pgfutil@tempdima
    \sbox\pgfutil@tempboxa{$\m@th\fam\pgf@math@fam#1#2$}%
    \advance\pgfutil@tempdima-\dp\pgfutil@tempboxa
    \pgfutil@tempdimb\wd\pgfutil@tempboxa
    \pgfpathmoveto{\pgfqpoint{0pt}{\pgfutil@tempdima}}%
    \pgfpathlineto{\pgfqpoint{-\pgfutil@tempdimb}{\pgfutil@tempdima}}%
    \pgfusepath{stroke}
    \ifpgf@depth
      \pgf@relevantforpicturesizetrue
      \pgfpathmoveto{\pgfqpoint{0pt}{-\pgfutil@tempdimb}}%
      \pgfusepath{use as bounding box}%
    \fi
  \endpgfpicture
}
\makeatother

\newcommand{\sscr}{\scriptscriptstyle\rm}

\usepackage{manfnt}
\reversemarginpar

\usepackage{mathrsfs}
\newcommand{\Dd}{{\mathscr{D}}}
\newcommand{\scri}{{\mathscr{I}}} 
 
\usepackage{tikz}
\usetikzlibrary{arrows, decorations.pathmorphing, positioning,calc}

\DeclarePairedDelimiter{\braket}{\langle}{\rangle}
\DeclarePairedDelimiter{\abs}{\lvert}{\rvert}
\newcommand\restr[2]{{
 \left.\kern-\nulldelimiterspace 
 #1 
  \vphantom{\big|} 
  \right|_{#2} 
  }}

\theoremstyle{plain}%

\newtheorem*{problem*}{Problem}

\theoremstyle{remark}%

\theoremstyle{definition}%

\newcommand{\vol}{\eps}
\newcommand{\bS}{S}

\begin{document}

\title{\bf Horizon dynamical entropies and phase space topology}

\author{\Large{Matteo Bruno and Simone Speziale}
\smallskip \\ 
\small{\it{Aix Marseille Univ., Univ. de Toulon, CNRS, CPT, UMR 7332, 13288 Marseille, France}} }
\date{\today}

\maketitle

\begin{abstract}
We study phase space topologies induced by infinite-dimensional norms as a means to solve the problem of `integrability' of Hamiltonian generators corresponding to diffeomorphisms of a null hypersurface, with the specific goal in mind to compare different prescriptions for the dynamical entropy of black holes, which replaces the area in non-stationary regimes. 
We first show that for a scalar field, there exists a family of norms that would provide integrability for different polarizations. We then show that a single member of this family applies to the gravitational field, and it is given by the Sobolev norm already considered in the literature. The obstruction to the other candidates lies in the semi-positivity of the metric.
\end{abstract}

\tableofcontents

\begin{flushright}
\emph{In memory of Jurek Lewandowski}
\end{flushright}

\section{Introduction}

A beautiful and well-known result about black hole physics is that the horizon's area, which is conjectured to provide a notion of black hole entropy \cite{Bekenstein:1972tm,Hawking:1976ra}, can be derived as a Noether charge \cite{Wald:1993nt}. This result has been recently extended to non-stationary situations \cite{Rignon-Bret:2023fjq,Hollands:2024vbe,Visser:2024pwz}, with the introduction of a suitable notion of \emph{dynamical entropy}, given by one quarter of the \emph{dynamical area}, the latter being the area of the dynamical (aka apparent) horizon sitting inside the expanding event horizon, see also \cite{Ashtekar:2025qqa}. The dynamical area is one of the infinitely many Noether charges that can be associated to the symmetry group of a generic null surface following \cite{Chandrasekaran:2018aop}. This construction provided useful background for Jurek's work on charges and multipoles of non-expanding horizons \cite{Ashtekar:2021wld,Ashtekar:2021kqj}. 
While Noether charges are notoriously ambiguous, a unique set was there identified using a prescription inspired by the treatment by Wald and Zoupas of null infinity \cite{Wald:1999wa}. In particular, the charges satisfy the crucial physical requirements of covariance and stationarity, which were taken as building block of the generalization of the Wald-Zoupas prescription in \cite{Odak:2022ndm}.

The interpretation of the new charge as dynamical entropy is however restricted to perturbations of Killing horizons. An immediate limitation of more general interpretations is that the charge does not vanish on a Minkowski light cone. A solution to this limitation was proposed in \cite{Odak:2023pga}. By relaxing the stationarity condition so that it holds for the symmetry flux as opposed to hold for the full symplectic potential, it is possible to find a 1-parameter family of covariant Noether charges. While a unique set is selected by the stronger stationarity condition of \cite{Chandrasekaran:2018aop},  an alternative, and also unique, set can be selected requiring all charges to vanish in Minkowski. These two  options can play useful and complementary roles, depending on the physical system that one wants to consider, and can be understood as `Dirichlet' and `conformal' polarizations, or insofar as the spin-0 sector is concerned, as momentum and position polarizations respectively, namely $p\d q$ vs $-q\d p$.

The ambiguity in the Noether charges can also be fixed in a  completely independent way, looking at the canonical generators of the symmetry in the phase space. With open boundary conditions (aka leaky, or dissipative), some symmetry transformations do not correspond to Hamiltonian vector fields, hence their generators do not exist, or in the language often used in this context, are `not integrable'. It was proposed in \cite{Ashtekar:1981bq,Ashtekar:2024stm} to deal with it introducing a norm in the phase space, and using the topology it induces to define the canonical generator extending by continuity the one defined in the integrable subset of the phase space. This procedure was successfully applied at null infinity, and the results there obtained are the same as with the Wald-Zoupas prescription, which is quite remarkable, given the different logic and requirements involved. In this paper, we investigate the topological prescription at finite distance, and compare it with the generalized Wald-Zoupas prescription. 

We first consider a 1-dimensional toy model of functions on a finite interval. We prove that the momentum polarization is selected by the phase space topology induced by the Sobolev norm, while the position polarization is selected using a boundary-improved Lebesgue norm, where the function's standard Lebesgue norm is supplemented by independent Lebesgue norms at the boundary values.  The norm consists of the standard \(L^2\) norm supplemented by terms controlling the point-wise behaviour of the derivative at the boundary, namely the `momentum'.
We then show that analogous results holds in the infinite-dimensional case of a scalar field on a null hypersurface with boundaries. In this case, the Sobolev result was already given in \cite{Ashtekar:2024stm}. Here we provide further details on the proof and its independence of the choice of inverse metric. For the converse case, we show that the natural extension of the 1-dimensional toy model applies straightforwardly. That is, the boundary point-wise control of the Sobolev norm is replaced by a norm with an \(L^2\) boundary term involving the derivative. It is therefore possible to select the Hamiltonian generator of diffeomorphisms for a scalar field on a generic null hypersurface (with dynamical metric and non-conservative boundary conditions for the scalar field) using either choices of topology. The resulting Hamiltonians differ by an integration by parts in field space. Interestingly, we find a  family of norms realizing integrability for mixed Robin-type polarizations.

The situation is substantially different for the gravitational field. For the metric, we find a natural norm that realizes the momentum polarization, whereas the norm providing the opposite construction for the scalar field fails to implement the desired properties. We identify the origin of this obstruction in the positivity of the metric: it is not sufficient for the metric to vanish at the corner, its first derivative must also be compatible with maintaining positive definiteness in the bulk. This requirement has no analogue for a scalar field, whose values are unrestricted in sign, but is intrinsic to the space of Riemannian metrics. As a consequence, only the momentum polarization of \cite{Chandrasekaran:2018aop} appears for the moment to be consistent with integrability from phase space topology.
The main themes of this paper -- quasi-local horizons and the covariant phase space -- have been deeply influenced by Jurek's work and by his profound and elegant approach to physics. He had a remarkable ability to combine physical questions and intuition with detailed and precise mathematical analysis. We have tried to carry this spirit into the present work, and we dedicate this paper to his memory.

\section{Charges on null hypersurfaces}

Noether's theorem provides a conserved current for any diffeomorphism,
\be
j_\xi := I_\xi\th -i_\xi L, \qquad dj_\xi\eqons 0.
\ee 
In this formula, we are using the notation of the covariant phase space as in \cite{Speziale:2025lkm}, where $L$ is the Lagrangian seen as a 4-form, and the symplectic potential current $\th$ is defined through the variation $\d L\eqons d\th$, where $\eqons$ means on-shell of the field equations.
Furthermore, the current is also exact on-shell, 
\be\label{jdq}
j_\xi = C_\xi +dq_\xi \eqons dq_\xi,
\ee 
where $C_\xi$ are the constraint part of the field equations.
As a consequence, integrating \eqref{jdq} over a 3-dimensional manifold with two boundaries $S_{1,2}$, we obtain a flux-balance law determining the variation of surface charges 
\be
Q_\xi[S_2]-Q_\xi[S_1] \eqons F_\xi := \int j_\xi, \qquad Q_\xi[S]=\oint_S q_\xi.
\ee

The Noether current is however ambiguous, and the freedom can be parametrized in terms of the equivalence class of symplectic potentials 
\be\label{th'}
\th\sim \th' = \th+\d\ell-d\vth.
\ee
Often, we are only interested in the symplectic potential on a specific boundary, and then this equivalence class only plays a role after pull-back.
The usefulness of the flux-balance laws relies on the possibility of identifying a preferred representative, possibly unique. A  successful prescription was proposed by Wald and Zoupas \cite{Wald:1999wa} for null infinity, extended in \cite{Chandrasekaran:2018aop} to null hypersurfaces, and generalized and systematized in \cite{Odak:2022ndm} for arbitrary boundary conditions, including corner terms and field-dependent diffeomorphisms. For the specific case of pull-back on a null hypersurface $\cN$, we can identify in the equivalence class \eqref{th'} a 1-parameter family of covariant symplectic potentials, given in units $16\pi G=1$ by \cite{Odak:2023pga,Chandrasekaran:2023vzb}
\begin{align}\label{Thc}
\th^{\sscr (c)} = \big( \s^{\m\n} \d \g_{\m\n}  +(2-c)\d\th \big)\eps_\cN + (1-c)\th\d\eps_\cN.
\end{align}
The notation used in this expression is as follows. We denote the null hypersurface's normal $l_\m$, with inaffinity $k$, expansion $\th$,\footnote{We use the same standard symbol for  the symplectic potential and the expansion, hoping that no confusion arises thanks to the context and the different tensorial nature.} and shear $\s_{\m\n}$. These are related to the induced metric $\g_{\m\n}$ on $\cN$ via $\pounds_l \g_{\m\n}=\s_{\m\n}+(\th/2)\g_{\m\n}$.
The volume 3-form can be defined from the spacetime volume form $\vol$ introducing an auxiliary null vector $n^\m$ satisfying $n^\m l_\m=-1$, often called `rigging vector', and it satisfies
\be
\eps_\cN = i_n \vol, \qquad i_l\eps_\cN=\eps_S, \qquad d\eps_S=\th\eps_\cN,
\ee
where $\eps_S$ is the area 2-form on the space-like cross sections of $\cN$. 
Following Chandrasekaran, Flanagan and Prabhu (CFP) \cite{Chandrasekaran:2018aop}, we restricted the variations in \eqref{Thc} to satisfy the boundary conditions $\d l_\m=\d l^\m=\d k=0$, which can be imposed for any metric by a suitable coordinate choice, without any restriction on the dynamics. 
With these boundary conditions, $c=0$ is the standard Einstein-Hilbert potential, and different values of $c$ describe different `polarizations' associated with adding a boundary Lagrangian.\footnote{We use the term polarization in a loose sense here, since not every boundary Lagrangian leads to a new polarization. To see the precise relation, consider for instance the 1-parameter family of polarizations $Q_\th=\cos\th q +\sin\th p$, $P_\th=-\sin\th \, q+\cos\th \, p$. Then 
\[
\th^{\sscr (c)} = (c-1)p\d q+(c-2)q\d p = P_\th \d Q_\th +\d\ell, \qquad \ell=\sin2\th (p^2-q^2), \qquad c=1+\cos^2\th.
\]
\label{FootPol}
}

Since the boundary conditions do not constrain the dynamics, they are `leaky' rather than conservative. When the null hypersurface is interpreted as incoming', they permit dissipation; when it is `outgoing', they instead permit amplification. The different polarizations of the symplectic potential can therefore be understood as corresponding to physical systems with different conservative sectors. From this perspective, the fact that different polarizations lead to different charges is analogous to the familiar thermodynamic distinction between internal energy and free energy: changing the boundary conditions from adiabatic to isothermal changes the appropriate thermodynamic potential, and hence the quantity that plays the role of the conserved energy.

The diffeomorphisms preserving the above boundary conditions form the CFP group. They are generated by tangent vector fields that can be parametrized by
\be
\xi=(T+\l W)\p_\l+Y^A\p_A,
\ee
where $\l$ is an affine parameter along the null geodesics, $x^A$ are the angular coordinates, and $T$, $W$ and $Y$ are arbitrary functions on the sphere (but time-independent), representing respectively super-translations, Weyl rescaling and diffeomorphisms of the cross sections. 
The surface charges associated to \eqref{Thc} can be obtained either `integrating the fluxes' $j_\xi^{\sscr (c)}=I_\xi\th^{\sscr (c)}$ \cite{Hopfmuller:2018fni,Odak:2023pga,Ciambelli:2023mir}, or using the improved Noether charge formula \cite{Odak:2023pga}, and the result is
\be\label{Qc}
Q^{\sscr (c)}_\xi[S] = 2\oint_S\left( W\left(1-\f c2\l\th\right) -T\th - Y^A\eta_A \right)\eps_S.
\ee

We can make the meaning of these charges more concrete considering some examples. If $\cN$ is the Kerr horizon, and we choose $\xi$ to be the horizon generator $\xi=\p_t+\Om_{\sscr \cH}\p_\phi\stackrel{\sscr\cH}=\k \l\p_\l$, then 
$Q_\xi[S] = 2 \k A$, and it coincides with the familiar result obtained using the Komar integral, regardless of the value of $c$. 
If $\cN$ is a non-expanding horizon, we can set $\th=0$, and also restrict $W$ to be constant and $Y$ to be a conformal Killing vector of the sphere, to preserve the universal structure of non-expanding horizons proposed in \cite{Ashtekar:2021wld}, and we recover the charge expressions derived in \cite{Ashtekar:2021kqj}.
If $\cN$ is a null cone in Minkowski, and we set $T=0$ so to preserve the tip of the cone, $Q_\xi[S] = 2(1-c)\oint W\eps_S$.

The parameter $c$ gives a family of different polarizations in the `spin-0 sector' with conjugate pair the volume and expansion of the hypersurface. Notable examples of this family are the `position polarization' (namely $q\d p$) for $c=1$, and the `momentum polarization' (namely $-p\d q$) for $c=2$.
Notice that $\th^{\sscr (c)}$ is independent of the gauge choices that go in the description of the null hypersurface geometry, specifically the class III transformations $(l,n)\to (A l, A^{-1}n)$, and class I transformations $(l,n,m)\to (l,n+a\bar m+\bar a m,m+a l)$.\footnote{The classification follows e.g. \cite{Chandra}, and it is based on the internal Lorentz transformations of the Newman-Penrose tetrad.}
This is what we mean by \emph{covariant} polarizations, and as argued in \cite{Wald:1999wa,Odak:2022ndm}, it is a crucial criterium to reduce the ambiguities in the Noether currents and symplectic potential. In particular, it guarantees that the Noether current provide a realization of the symmetry algebra in the phase space \cite{Rignon-Bret:2024wlu}.

The second crucial criterium is stationarity \cite{Wald:1999wa,Odak:2022ndm}. In its stronger version, it requires that the chosen potential vanishes for arbitrary variations around solutions suitably identified as `stationary'. In the case of null hypersurfaces, a useful notion of stationarity is to be shear and expansion free, and we will refer to such hypersurfaces as \emph{non-expanding horizons} (NEH) \cite{Ashtekar:2021wld}, with a slight abuse of language.\footnote{Non-expanding horizons coincide with shear and expansion free null hypersurfaces in vacuum, and a slight subset thereof in the presence of matter.} Inspecting \eqref{Thc} we see that implementing stationarity fixes $c=2$, thus singling out a unique representative. This is the Wald-Zoupas potential identified in  \cite{Chandrasekaran:2018aop}. A weaker version of the stationarity criterium, suggested in this context in \cite{Odak:2023pga}, is to require the vanishing of the Noether flux $I_\xi{\th}$, as opposed to the full potential $\th$, on the stationary solutions, which we keep to be the NEHs. Since the symmetry group preserves the expansion, the condition is satisfied for any value of $c$. This weaker version thus not remove the ambiguity by itself, and this feature allows us to add a additional requirement: the vanishing of all charges in Minkowski. Inspection of \eqref{Qc}, and recalling that $\th=2/\l$ for an outgoing null cone in Minkowski, shows that the weaker criterium and the new requirement lead to the unique choice $c=1$. Summarizing, we have a 1-parameter family of covariant options, with two special values corresponding to the momentum and position polarizations in the spin-0 sector, and singled out by two different ways of interpreting the stationarity condition: vanishing of the full potential, and non-zero charges in Minkowski, or vanishing of the symmetry flux only, and all charges vanishing in Minkowski.

\subsection{Hamiltonian generators}\label{SecHamGen}

In standard Hamiltonian systems, the Noether charge is also the canonical generator of the symmetry transformation in phase space. This is not necessarily the case with leaky or dissipative boundary conditions, because some of the symmetry transformations `see' the dissipation and are not Hamiltonian vector fields, see discussion in \cite{Speziale:2025lkm}. This problem is often referred to as `non-integrability' of the charges in the literature, and shows up explicitly in our context.
The symplectic 2-form obtained from \eqref{Thc} is 
\be
\om=\d \th^{\sscr (c)} = \d (\s^{\m\n}\eps_\cN) \cw \d \g_{\m\n}  - \d\th \cw \d\eps_\cN,
\ee
independently of $c$. The polarizations play however a key role in determining the integrability of the charges. We have
\begin{align}
-I_\xi\om &= - \d_\xi(\s^{\m\n}\eps_\cN)\d\g_{\m\n} +\d(\s^{\m\n}\eps_\cN)\d_\xi\g_{\m\n} + \d_\xi\th\d\eps_\cN - \d\th\d_\xi\eps_\cN
\nn\\&=
\d(\s^{\m\n}\eps_\cN \pounds_\xi \g_{\m\n}) - d\left( \s^{\m\n}i_\xi \eps_\cN\d\g_{\m\n}\right) - \d(\th\d_\xi\eps_\cN) + d(\th i_\xi\d \eps_\cN).
\end{align}
The $\d$ exact terms are the `integrable' ones, and we see that the obstruction to integrability is a corner contribution only, in agreement with the general result \cite{Iyer:1994ys}.
The `integrable/non-integrable' split in the spin-2 sector involving the shear is unambiguous, but the spin-0 sector can be arbitrarily mixed up using the identity
$\d\d_\xi(\th\eps_\cN)=d\d(\th i_\xi\eps_\cN)$, leading to a 1-parameter family of prescriptions for the generator:
\begin{align}\label{Ixic}
&-I_\xi\om = \d j^{\sscr (c)}_\xi - d X_\xi,\qquad
 j^{\sscr (c)}_\xi = \s^{\m\n}\eps_\cN \pounds_\xi \g_{\m\n} +(2-c) \d_\xi\th \eps_\cN +(1-c) \th\d_\xi\eps_\cN, 
\end{align}
where
\be
X_\xi = \big(\s^{\m\n}\d\g_{\m\n} +(2-c) \d \th\big) i_\xi\eps_\cN +(1-c)\th i_\xi\d\eps_\cN
\ee
is the dissipation term causing the obstruction to integrability.

This result shows that there are two special values of $c$, which notably corresponds precisely to the two proposals reviewed above for the stationarity requirement. In both cases, integrability requires the shear to vanish at the corners of the null hypersurface. For $c=2$, we also need the expansion to vanish. For $c=1$, we need the stronger boundary condition $\d\th=0$ at the corner, or the area 2-form to vanish, since $i_\xi\eps_\cN=f\eps_S+i_Y\eps_\cN$ and only the first term contributes to the pull-back on a space-like corner.

The Hamiltonian generator offers also an independent perspective on the question of selecting $c$. 
It was proposed in \cite{Ashtekar:1981bq,Ashtekar:2024stm} that a prescription for integrability can be given from the topology of the phase space. If the configurations with a specific choice of $c$ are dense, and the others are not, then it is natural to choose that polarization, and define the generator in the full phase space extending the corresponding Noether charge by continuity. This strategy has been used successfully at null infinity, and we investigate here its application to a physical null hypersurface.

To that end, observe that the field equations alone do not induce a topology on the covariant phase space. We can however provide one introducing a norm. The question is then whether there exists a norm such that the `extension by continuity' argument selects any special values of $c$. Specifically to our question on the dynamical entropy, 
the momentum polarization $c=2$ can be selected by a topology in which vanishing derivatives of a function at the boundary form a dense subset but vanishing values of the function do not, and the position polarization $c=1$ by the converse, namely a topology in which vanishing values of a function at the boundary form a dense subset but vanishing derivatives of the function do not. Whether norms inducing such topologies exists requires a careful mathematical analysis, which is the object of the following Sections. We also remark that careful considerations on the topology of the phase space also played a key role in the recent construction of gravitational observables at null infinity \cite{AndradeeSilva:2026mpa}.

The result is that we were only able to find a norm that selects the $c=2$ momentum polarization, while a norm that selects the position polarization in the simpler case of a scalar field on a null hypersurface fails to apply to the gravitational case because of the requirement of semi-positiveness of the metric.

\section{Phase space norms}

Before proving the new results, we review some basic aspects of Lebesgue and Sobolev norms and their associated spaces. We use a simple one-dimensional case to illustrate how choosing carefully the norm we control whether vanishing derivatives of a function at the boundary form a dense subset but vanishing values of the function do not, or the converse. We then apply the same construction to a scalar field theory on a bounded null hypersurface in curved spacetime. In this case we find again that both options are possible. The first option is selected by the Sobolev norm proposed in \cite{Ashtekar:2024stm}, and our analysis provides more background material and more details to support the result used there. The second option is selected by a new norm, which replaces point-wise control with a Lebesgue norm on the boundary derivatives.
Finally, we move to the metric case, where applying the same construction we will find that only the first option seems possible, and point out the obstructions that should be overcome in order to be able to implement the second option.

\subsection{Lebesgue and Sobolev spaces}
The two fundamental spaces in functional analysis considered in this paper are the Lebesgue space \(L^2\) and the Sobolev space \(H^1\). They are both Hilbert spaces, and, on a compact interval \(I\), they can be defined as the completions of the space of smooth functions \(C^\infty(I)\) with respect to two different norms. The first is the \(L^2\) norm,
\begin{equation}
\norm{f}_{L^2}^2=\int_I \abs{f(x)}^2 dx,
\end{equation}
while the second is the Sobolev norm,
\begin{equation}
\norm{f}_{H^1}^2=\int_I\left(\abs{f}^2+\abs{\dot{f}}^2\right) dx,
\end{equation}
where the dot denotes the derivative with respect to \(x\).

The \(L^2\) completion loses, in general, all pointwise information about a function and, in particular, does not control its values at the boundary, nor those of the derivative. A standard example is provided by the sequence of characteristic functions 
\[\chi_{\left[\frac{1}{n},1-\frac{1}{n}\right]}(x)=\begin{cases}
    1 & \frac{1}{n}\leq x\leq 1-\frac{1}{n},\\
    0 & \text{otherwise,}
\end{cases}\]
which converges in the \(L^2\) norm on \([0,1]\) to the constant function \(1\), despite vanishing at the boundary for any finite $n$. In contrast, the Sobolev space retains additional information about the behavior of its elements and, in particular, controls the boundary values. Specifically to one dimension, there is an even stronger result: every \(H^1\) function admits an absolutely continuous representative, and its weak derivative\footnote{Namely defined in the sense of distributions.} is represented by an \(L^2\) function. Thus, although elements of \(H^1(I)\) are still equivalence classes of functions, they possess a representative for which pointwise values are well defined and which, in particular, has well-defined boundary values.

Notice that the distinction between the two spaces is not in the functions they contain, but rather in the topology and the additional regularity controlled by their respective norms.
 There is in fact a natural continuous embedding
\begin{equation}
H^1(I)\hookrightarrow L^2(I),
\end{equation}
since
\begin{equation}
H^1(I) = \left\{f\in L^2(I)\ \middle|\ \dot f\in L^2(I)\right\},
\end{equation}
where the derivative is understood in the weak sense. Moreover, \(H^1(I)\) is dense in \(L^2(I)\) with respect to the \(L^2\) norm.\footnote{For every point in $L^2(I)$, there exists a sequence in the Sobolev space converging to that point in $L^2$ norm. This implies that the completion of $H^1(I)$ with respect the $L^2$ norm is $L^2(I)$.}

The constructions generalize to higher-dimensional manifolds and to non-compact domains, with suitable modifications. For the purposes of this work, it is useful to formulate them for sections of a vector bundle, which we denote generically $u$. Let \(E\to M\) be a Riemannian vector bundle over a compact Riemannian manifold \(M\), equipped with the volume form $\eps$ induced by the metric, and let \(\Gamma(E)\) denote the space of smooth sections of \(E\). By choosing a connection \(D\) on \(E\), one can define the \(L^2\) and \(H^1\) norms by
\begin{subequations}
\begin{equation}
\norm{u}_{L^2}^2=\int_M \abs{u}^2 \vol,
\end{equation}
\begin{equation}\label{Sobolev}
\norm{u}_{H^1}^2=\int_M\left(\abs{u}^2+\abs{Du}^2\right) \vol.
\end{equation}
\end{subequations}
Here, \(\abs{\cdot}\) denotes the pointwise norm induced by the Riemannian metric on \(M\) and the inner product on the fibers of \(E\); in particular, it is the norm on \(E\) when applied to \(u\), and the induced norm on \(T^*M\otimes E\) when applied to \(Du\). The corresponding spaces \(L^2(M,E)\) and \(H^1(M,E)\) are defined as the completions of \(\Gamma(E)\) with respect to these norms.

If $M$ is compact, different choices of Riemannian metric and connection give rise to equivalent Sobolev spaces \cite{Hebey_1996}; namely, they consist of the same elements and are isomorphic as topological vector spaces. More precisely, if \(\|\cdot\|_1\) and \(\|\cdot\|_2\) are the norms induced on \(\Gamma(E)\) by two different choices of metric and connection, then there exist constants \(\alpha,\beta>0\) such that
\[
\alpha\|u\|_1\leq \|u\|_2\leq \beta\|u\|_1
\]
for every \(u\in\Gamma(E)\). In other words, the two norms are equivalent and therefore induce the same topology. This freedom will be useful below, since it allows us to choose a convenient auxiliary Riemannian metric when no canonical metric is available. Notice, however, that in dimensions greater than one the \(H^1\) norm does not identify a unique continuous representative, hence fails to generally control pointwise behavior: in particular, the Sobolev embedding \(H^1(M)\hookrightarrow C^0(M)\) fails when \(\dim M>1\). Thus, \(H^1\) retains more information than \(L^2\), but it does not in general provide pointwise control. Nonetheless, pointwise control is retained for \emph{boundary} values, because the map that associates to each function its restricion on the boundary is well-defined and continuous. As a consequence, functions that vanish at the boundary form a closed subset and are not dense. This boundary control is crucial for our discussion on polarizations, and it can be proved using the trace operator.

Let \(M\) be a compact Riemannian manifold with boundary. For smooth functions, the \emph{trace operator} is defined by
\begin{equation}
\begin{matrix}
T: & C^{\infty}(M) & \longrightarrow & C^{\infty}(\partial M)\,;\\
& f &\longmapsto & \restr{f}{\partial M}\,.
\end{matrix}
\end{equation}
The trace theorem states that this operator extends uniquely to a bounded operator
\begin{equation}
T:H^1(M)\longrightarrow L^2(\partial M).
\end{equation}
Consequently, the space of \(H^1\) functions with vanishing trace,
\[  
H^1_0(M):=\ker T,
\]
is a closed subspace of \(H^1(M)\). Notice that, for \(H^1\) functions, ``vanishing at the boundary'' is understood in terms of the trace rather than pointwise evaluation.

The boundedness of the trace operator can be proven easily in the present setting. We give here the proof for pedagogical reasons and will also be useful in the remainder of the work. Let \(n\) be the outward-pointing unit normal to \(\partial M\), so that the induced volume form on the boundary is
\[ \epsilon_{\partial M}=i_n \vol,\]
where \(i_n\) denotes the interior product. Let \(\tilde n\) be any smooth extension of \(n\) to the whole of \(M\).  Then
\begin{equation}\label{Tbounded}
\begin{split}
\int_{\partial M}\abs{f}^2 i_n \vol
&=\int_{\partial M}\abs{f}^2 i_{\tilde n}\vol\\
&=\int_M d\left(\abs{f}^2 i_{\tilde n}\vol\right)\\
&=\int_M\left(2fdf\wedge i_{\tilde n}\vol+\abs{f}^2d(i_{\tilde n}\vol)\right)\\
&=\int_M\left(2f\,i_{\tilde n}df+\abs{f}^2\operatorname{div}\tilde n\right)\vol\\
&\leq\int_M\abs{f}^2\vol+\int_M\abs{i_{\tilde n}df}^2\vol+\int_M\abs{f}^2\operatorname{div}\tilde n\,\vol\\
&\leq\int_M\abs{f}^2\vol+\int_M\abs{\tilde n}^2\abs{\nabla f}^2\vol+\int_M\abs{f}^2\operatorname{div}\tilde n\,\vol\\
&\leq\left(1+\max_M\abs{\operatorname{div}\tilde n}\right)\int_M\abs{f}^2\vol+\max_M\abs{\tilde n}^2\int_M\abs{\nabla f}^2\vol\\
&\leq \max\left(1+\max_M\abs{\operatorname{div}\tilde n},\max_M\abs{\tilde n}^2\right)\int_M\left(\abs{f}^2+\abs{\nabla f}^2\right)\vol.
\end{split}
\end{equation}
In  the second equality above, we used Stokes' theorem for the second equality. For the fourth equality, we used
\[
d(i_{\tilde n}\vol)=(\operatorname{div}\tilde n)\vol,
\]
and the identity
\[ (i_{\tilde n}df)\vol - df\wedge i_{\tilde n}\vol=i_{\tilde n}(df\wedge \vol)=0.\]
The first inequality follows by recognizing
\[ \int_M f\,i_{\tilde n}df\,\vol\]
as an \(L^2\) inner product, and using the elementary estimate $2\braket{x,y}\leq\abs{x}^2+\abs{y}^2.$ For the second inequality, we used
\[\abs{i_{\tilde n}df}^2=\abs{g(\tilde n,\nabla f)}^2 \leq\abs{\tilde n}^2\abs{\nabla f}^2,\]
where \(\nabla\) is the gradient with respect to the Riemannian metric \(g\).
Finally, since \(M\) is compact, the max in the last inequality is bounded. There exists thus a constant \(C>0\), depending only on the chosen extension \(\tilde n\) and the Riemannian geometry of \(M\), such that
\begin{equation}
\int_{\partial M}\abs{f}^2 i_n \vol
\leq
C
\int_M
\left(
\abs{f}^2+\abs{\nabla f}^2
\right)\vol.
\end{equation}
This is the desired estimate for the trace operator, proving that it is bounded.

The discussion above highlights a distinction that will be important in what follows. The \(L^2\) topology controls bulk square-integrability but does not retain boundary information, whereas the \(H^1\) topology controls one additional derivative and admits a continuous trace to the boundary. Consequently, when constructing Hilbert spaces associated with fields or geometries on a manifold with boundary, the choice of norm determines which boundary data survive in the completion. In particular, one can combine bulk \(H^1\) control with an explicit boundary \(L^2\) contribution when it is necessary to retain boundary information that would otherwise be lost in the completion. This will be the relevant construction for the null hypersurface considered in the following sections.

\subsection{One-dimensional case}

The one-dimensional toy model of the integrability problem is to consider smooth functions on the interval $[0,1]$, and equip this space with a symplectic form 
\be
\Om=\int_0^1\d\dot f \cw \d f\, dx,
\ee
and then evaluate the canonical generator of line diffeomorphisms $x\to x'=x+\xi(x)$:
\be
-I_\xi\Om=\d\int_0^1\dot f \pounds_\xi f\, dx - \xi\dot f \d f\Big|^1_0 
=-\d\int_0^1f \d_\xi\dot f\, dx+ \xi f \d\dot f\, \Big|^1_0 .
\ee
The two options represent the momentum and position polarization respectively.\footnote{The notation is $\d_\xi f=\pounds_\xi f$ and $\d_\xi \dot f = \pounds_{\p_x}\pounds_\xi f=\pounds_\xi \dot f+\dot\xi\dot f$, and we used $\dot\xi dx=\pounds_\xi dx.$}

Consider then the following spaces:
\begin{align}
& Q=\{f\in C^\infty([0,1])\mid f(0)=f(1)=0\},\\
& P=\{f\in C^\infty([0,1])\mid \dot{f}(0)=\dot{f}(1)=0\}.
\end{align}
The $L^2$ norm makes both subsets dense, in other words the topology it induces is too weak to pick a preferred polarization.
If we find a norm such that $P$ is dense and $Q$ is not, this will correspond to selecting the momentum polarization, and conversely. Hence the labels chosen.
We have already proved in the previous section that the Sobolev norm \eqref{Sobolev} and associated space $H^1([0,1])$ realize the first option. In this subsection we construct a norm that realizes the converse option, namely selecting the position polarization.

To find the desired norm, it is sufficient to work in the slightly broader space of continuous functions on $[0,1]$ that are $C^1$ in a neighborhood of the boundary. We define
\begin{equation}\label{WEnorm}
    \norm{f}^2
    \coloneqq \abs{\dot{f}(0)}^2
    +\abs{\dot{f}(1)}^2+
    \int_0^1 \abs{f(x)}^2\,dx.
\end{equation}
This norm satisfies the parallelogram identity and therefore defines, through the polarization identity, a scalar product $\braket{\cdot,\cdot}$. Thus, we may define a Hilbert space as the completion
\begin{equation}\label{Pspace}
    \overline{C^\infty([0,1])}^{\scriptscriptstyle\norm{\cdot}}.
\end{equation}
We now prove that this norm makes $P$ not dense, while $Q$ is dense.

The strategy to prove that the space of functions whose derivative vanishes at the boundary is not dense consists in considering the \emph{differential trace} operator, and showing that it is bounded and has a non-trivial kernel. 
We define the differential trace operator as
\begin{equation}
\begin{matrix}
    DT:&C^\infty([0,1])&\longrightarrow&\mathbb{R}^2\,;\\
    &f&\longmapsto&(\dot{f}(0),\dot{f}(1))\,.
\end{matrix}
\end{equation}
We equip  $\R^2$ with the standard Euclidean norm $|\cdot|$. This operator is bounded with respect to $\norm{\cdot}$ since
\begin{equation}\begin{split}
    \abs{DT(f)}^2
    &=
    \abs{\dot{f}(0)}^2+\abs{\dot{f}(1)}^2\\
    &\leq
    \int_0^1\abs{f(x)}^2\,dx
    +\abs{\dot{f}(0)}^2
    +\abs{\dot{f}(1)}^2
    =\norm{f}^2.
\end{split}
\end{equation}
In particular, $DT$ extends uniquely by continuity to the completion. Moreover,
\begin{equation}
    P=\ker DT,
\end{equation}
and therefore $P$ is closed by continuity. In particular, the space of functions whose derivative vanishes at the boundary cannot be dense.

Next, to prove that the space of functions vanishing at the boundary is dense, we need to characterize the Hilbert space \eqref{Pspace} more explicitly. To that end, we consider the linear map
\begin{equation}
\begin{matrix}
    U:&C^\infty([0,1])&\longrightarrow&\mathbb{R}^2\oplus L^2([0,1])\,;\\
    &f&\longmapsto&(\dot{f}(0),\dot{f}(1),f)\,.
\end{matrix}
\end{equation}
Notice that there is a slight abuse of notation here: the last component on the right-hand side denotes the image of the smooth function $f$ under the canonical embedding of $C^\infty([0,1])$ into $L^2([0,1])$, namely its equivalence class modulo equality almost everywhere.
Our strategy is to first establish that this map is a unitary isomorphism, and then that the space of functions vanishing at the boundary is dense on the image of this map.

The map $U$ is clearly an isometry, since we equipped $\R^2$ with the Euclidean norm, which is what we use for the boundary values in \eqref{WEnorm}. We shall prove that its image is dense in $\mathbb{R}^2\oplus L^2([0,1])$, from which it follows that $U$ extends to a unitary isomorphism between $\overline{C^\infty([0,1])}^{\scriptscriptstyle\norm{\cdot}}$ and $\mathbb{R}^2\oplus L^2([0,1]).$ Let
\[
(a,b,f)\in \mathbb{R}^2\oplus L^2([0,1]).
\]
Since $C^\infty([0,1])$ is dense in $L^2([0,1])$, there exists a sequence $\{f_n\}_{n\in\mathbb{N}}$ of smooth functions such that
\[
f_n\longrightarrow f
\]
in the $L^2$ norm. We now modify this sequence by introducing a new sequence $\{\tilde f_n\}_{n\in\mathbb{N}}$ in order to control the boundary values. To that end, it is simpler to work with the slightly broader space of $C^1$ functions in a neighbourhood of the boundary.\footnote{Alternatively, the sequences below can be replaced by smooth interpolations without affecting the argument.} Within that space, let $\tilde f_1=f_1$, and for $n\geq2$ define
\begin{equation}
    \tilde f_n(x)=
    \begin{cases}
        a\left(x-\frac1n\right)+f_n\left(\frac1n\right),
        &0\leq x<\frac1n,\\[2mm]
        f_n(x),
        &\frac1n\leq x\leq1-\frac1n,\\[2mm]
        b\left(x-1+\frac1n\right)+f_n\left(1-\frac1n\right),
        &1-\frac1n<x\leq1.
    \end{cases}
\end{equation}
By construction,
\[
U\tilde f_n
=
\left(
\dot{\tilde{f}}_n(0)=a,
\dot{\tilde{f}}_n(1)=b,
\tilde{f}_n
\right),
\qquad
\forall n\geq2.
\]

The convergence is established as follows.\footnote{In the following estimate we repeatedly use the elementary inequality $\abs{x+y}^2\leq2(\abs{x}^2+\abs{y}^2)$.}
\begin{align*}
    \norm{(a,b,f)-U\tilde f_n}^2
    &=
    \norm{f-\tilde f_n}_{L^2}^2\\
    &=
    \int_0^{\frac1n}
    \abs{
        f(x)
        -a\left(x-\frac1n\right)
        -f_n\left(\frac1n\right)
    }^2dx\\
    &\quad+
    \int_{\frac1n}^{1-\frac1n}
    \abs{f(x)-f_n(x)}^2dx\\
    &\quad+
    \int_{1-\frac1n}^1
    \abs{
        f(x)
        -b\left(x-1+\frac1n\right)
        -f_n\left(1-\frac1n\right)
    }^2dx\\
    &\leq
    2\int_0^1\abs{f(x)-f_n(x)}^2dx\\
    &\quad+
    2\int_0^{\frac1n}
    \abs{
        f_n(x)
        -a\left(x-\frac1n\right)
        -f_n\left(\frac1n\right)
    }^2dx\\
    &\quad+
    2\int_{1-\frac1n}^1
    \abs{
        f_n(x)
        -b\left(x-1+\frac1n\right)
        -f_n\left(1-\frac1n\right)
    }^2dx\\
    &\leq
    2\norm{f-f_n}_{L^2}^2\\
    &\quad+
    4\int_0^{\frac1n}
    \abs{
        f_n(x)-f_n\left(\frac1n\right)
    }^2dx
    +
    4\int_0^{\frac1n}
    \abs{
        a\left(x-\frac1n\right)
    }^2dx\\
    &\quad+
    4\int_{1-\frac1n}^1
    \abs{
        f_n(x)-f_n\left(1-\frac1n\right)
    }^2dx
    +
    4\int_{1-\frac1n}^1
    \abs{
        b\left(x-1+\frac1n\right)
    }^2dx\\
    &\leq
    2\norm{f-f_n}_{L^2}^2\\
    &\quad+
    4\int_0^{\frac1n}
    \abs{
        f_n(x)-f_n\left(\frac1n\right)
    }^2dx\\
    &\quad+
    4\int_{1-\frac1n}^1
    \abs{
        f_n(x)-f_n\left(1-\frac1n\right)
    }^2dx\\
    &\quad+
    \frac{4a^2}{n^3}
    +\frac{4b^2}{n^3}.
\end{align*}

For $n\to+\infty$, every term tends to zero:
\begin{enumerate}
    \item
    $\norm{f-f_n}_{L^2}\to0$ by construction;

    \item
    \[
    \int_0^{\frac1n}
    \abs{
        f_n(x)-f_n\left(\frac1n\right)
    }^2dx
    \longrightarrow0,
    \]
    and similarly
    \[
    \int_{1-\frac1n}^1
    \abs{
        f_n(x)-f_n\left(1-\frac1n\right)
    }^2dx
    \longrightarrow0,
    \]
    because, by continuity of $f_n$, the integrands vanish at the limiting boundary point while the integration domains shrink to zero; in particular, these terms are $o(1/n)$;

    \item
    \[
    \frac{4a^2}{n^3}
    +\frac{4b^2}{n^3}
    \longrightarrow0.
    \]
\end{enumerate}

Therefore,
\[
U\tilde f_n\longrightarrow(a,b,f).
\]
Since $(a,b,f)$ was arbitrary, the range of $U$ is dense in
\[
\mathbb{R}^2\oplus L^2([0,1]).
\]

This characterization immediately allows us to prove the first requirement, namely that the functions vanishing at the boundary are dense. Indeed, since the space of smooth functions vanishing at the boundary is dense in $L^2([0,1])$, for every
\[
(a,b,f)\in \mathbb{R}^2\oplus L^2([0,1])
\]
we may choose a sequence $\{f_n\}$ of smooth functions such that $f_n(0)=f_n(1)=0$ and $f_n\longrightarrow f$ in $L^2([0,1])$. The sequence
\[
(a,b,f_n)
\]
then converges to $(a,b,f)$ in $\mathbb{R}^2\oplus L^2([0,1])$. Mapping it back through the inverse of the unitary extension of $U$, we obtain a sequence converging in the norm $\norm{\cdot}$ whose elements vanish at the boundary and whose derivative traces coincide with those of the function being approximated. Hence the space $V$ is dense in $\overline{C^\infty([0,1])}^{\scriptscriptstyle\norm{\cdot}}.$

\section{Scalar field on null hypersurface}
Let us now consider a regular portion $\cN$ of a null hypersurface, with topology
\be\label{NIS}
\cN\simeq I\times\bS,
\ee
where $\bS$ is a 2-sphere, and $I$ a compact interval along the null directions. The latter are geodesics of the ambient spacetime, and we can always choose to parametrize them with an affine parameter $s$ such that $I=[s_0,s_1]$. This choice eliminates all freedom of affine reparametrizations.\footnote{Given an arbitrary affine parameter $s$, the two boundaries will generically be at non-constant values $s_i=f_i(\th,\phi)$, and the freedom of affine reparametrizations is $s\to s'=A(\th,\phi)s+B(\th,\phi)$. Choosing $A=(s_1-s_0)/(f_1-f_0)$ and $B=s_0-Af_0$, we exhaust the freedom and set the boundaries at constant values $s_i$ of the new parameter.}
This manifold has a natural fiber bundle structure, with $S$ the base, and null geodesics the fibers. The intrinsic metric $q$ is degenerate, more precisely positive-semidefinite, and we denote $l$ its null eigenvector, namely the tangent to the null geodetics. This defines a map between bundles
\be
\begin{matrix}
q: &T\cN &\longrightarrow & T^*\cN,\\
& X &\longmapsto& q(X,\cdot),
\end{matrix}
\ee
whose kernel is $l$ and the image of each fiber $B_x\coloneqq q(T_x\cN)$ is two-dimensional. We can furthermore always introduce a horizontal split of the fiber bundle picking any 1-form $n$ transverse to $l$, namely such that $i_l n=-1$. This provides an Ehresmann connection for the bundle $\cN\to S$, namely a rank-2 distribution $H$ in $T\cN$, defined by $H=\ker n$, namely $n|_{H_x}=0$. Then,
\be\label{TNdec}
T_x\cN=H_x\oplus\operatorname{Span}(l(x)).
\ee
The rigging 1-form provides us with a preferred choice of (generalized) inverse metric $q^{-1}$, namely the one annihilating it, and the associated horizontal split for the cotangent bundle reads
\be\label{T*Ndec}
T_x^*\cN=B_x\oplus\ker q^{-1}_x = B_x\oplus\operatorname{Span}(n_x),
\ee
for every $x\in\cN$.
Both decompositions \eqref{T*Ndec} and \eqref{TNdec} are non-canonical, but unique once the rigging is chosen. It is often convenient to choose a rigging that is hypersurface orthogonal within $\cN$; it is then equivalent to choosing a foliation of $\cN$. Among these choices, it is always possible to choose a rigging that is Lie dragged by $l$. With these choices and $l=\p_s$ affinely parametrized, we have $n=-ds$ and 
$\eps_\cN=ds\w\eps_S$. In this coordinate system, the `rigging inverse' $q^{-1}$ chosen can be recognized as the Moore-Penrose inverse. The equivalence class of generalized inverses is then generated by 
\begin{equation}\label{qinvEC}
q^{-1}\sim q^{-1} + l\otimes X+X\otimes l+f\,l\otimes l,
\end{equation}
where $f\in C^\infty(\cN)$, and \(X(x)\in H_x\) for every \(x\in\cN\), and the positive semidefiniteness of the generalized inverse is ensured by the condition \(f\geq q(X,X)\geq0\).
The restriction of \(q\) to \(H_x\) is positive definite.

We consider a smooth scalar field on $\cN$, with symplectic structure
\be
\Om=\int_\cN\d\phi\cw\d\dot\phi\, \eps_\cN,
\ee
where we used the short-hand notation $\dot\phi:=i_ld\phi$. The symplectic flow of diffeomorphisms tangent to $\cN$ is  
\be
-I_\xi\Om=\d\int_\cN\dot\phi\pounds_\xi\phi\, \eps_\cN -\oint_{\p\cN} f\dot\phi\d\phi\, \eps_S
=-\d\int_\cN\phi\d_\xi\dot\phi\, \eps_\cN + \oint_{\p\cN} f\phi\d\dot\phi\, \eps_S.
\ee
We see again the problem of `non-integrability', as in Section ~\ref{SecHamGen}, associated with the fact that the flow does not preserve the symplectic structure, hence it is not Hamiltonian. The two equalities select two candidate Hamiltonians respectively in the `momentum polarization' and `position polarization', in the sense that the integrable piece is in the form $p\d_\xi q$ or $q\d_\xi p$. We define the subspaces 
\begin{align}
& Q=\{\phi\in C^\infty(\cN)\mid \phi|_{\p\cN}=0\},\\
& P=\{\phi\in C^\infty(\cN)\mid \dot{\phi}|_{\p\cN}=0\}.
\end{align}
As in the 1-dimensional case, an $L^2$ norm makes both subsets dense, hence the topology it induces is too weak to select a Hamiltonian.
A norm such that P is dense and Q is not will select the Hamiltonian in the momentum polarization, and viceversa. As we now prove, the first option is realized by the Sobolev norm. For the converse case, we construct a new norm with two components: one over the bulk of the null hypersurface, $L^2$ in the field and its angular derivatives; and one localised at the boundaries, $L^2$ in the field derivatives.  While the Sobolev norm can be understood as an energy for the scalar field, the new norm is a purely potential energy in the bulk, and a purely kinetic energy at the boundary.
Using these norms, either Hamiltonians can be defined in the full phase space, extending them by continuity from their integrable definitions in the dense subsets with respect to the chosen norm. 

We can also consider mixed Robin-type `polarizations', 
\begin{align}\label{Ixiphic}
-I_\xi\om&=\d j^{\sscr (c)}_\xi - dX_\xi, \qquad j_\xi^{\sscr (c)} = (c-2)\phi\d_\xi\dot\phi\, \eps_\cN+(c-1)\dot\phi\pounds_\xi\phi\, \eps_\cN, \\
& \hspace{3cm} X_\xi =(c-2)f\phi\d\dot\phi\, \eps_S + (c-1)f\dot\phi\d\phi\, \eps_S.
\end{align}
This is the scalar field analogue of the one-parameter family \eqref{Ixic}. We will comment on these at the end.

\subsection{Sobolev case}
The case of a scalar field on a null hypersurface in the momentum polarization, for which \(\dot{\phi}=0\) is dense, is analogous to the case of a scalar field at future null infinity treated in \cite{Ashtekar:2024stm}. In this Section we review the statement and provide explicit proofs. We cnsider the first Solbolev norm
\begin{equation}
    \norm{\phi}^2_{H^1}=\int_\cN\left(\abs{i_l d\phi}^2+\abs{d\phi}^2+\abs{\phi}^2\right)\epsilon_\cN.
\end{equation}
Here, the point-wise norm \(\abs{d\phi}\) is induced by an inner product on \(T^*\cN\) defined by the generalized inverse \(q^{-1}\) of the degenerate metric \(q\). Notice that, for any $\lambda_1,\lambda_2\in B_x\subset T_x\cN$ there exists $v_1,v_2\in T\cN$ such that $\lambda_i=q(v_i)$, moreover
\[  q^{-1}(\lambda_1,\lambda_2)=q(v_1,v_2)  .\]

The topology induced by this norm is independent of the choice of generalized inverse, namely it is the same along the equivalence class of generalized inverses. To prove this, we prove that the norm associated with any such inverse is equivalent to the one defined by the inverse \(q^-\). Using an equivalent generalized inverse which can be expressed as
\begin{equation}
    q^{-1}+l\otimes X+X\otimes l+f\,l\otimes l,
\end{equation}
the norm becomes
\begin{equation}\label{eq:norm_Sob-scalar}
    \norm{\phi}^2_{H^1} =\int_\cN\Big((f+1)\abs{i_l d\phi}^2+q^{-1}(d\phi,d\phi)+2\,i_l d\phi\,i_X d\phi+\abs{\phi}^2\Big)\epsilon_\cN.
\end{equation}
Using the inequality
\[ 2\,i_l d\phi\,i_X d\phi \leq \abs{i_l d\phi}^2+\abs{i_X d\phi}^2,\]
we define
\begin{equation}
    m=\max_{x\in\cN} \max_{\substack{\lambda\in B_x\\ q^{-1}(\lambda,\lambda)=1}} \abs{i_{X(x)}\lambda}^2.
\end{equation}
Every element of \(T_x^*\cN\) can be written as \(c_1\lambda+c_2 n_x\), for some \(c_1,c_2\in\mathbb R\) and some \(\lambda\in B_x\) satisfying \(q^{-1}(\lambda,\lambda)=1\). Consequently, we obtain
\begin{equation}
\begin{aligned}
    \abs{i_{X(x)}(c_1\lambda+c_2 n_x)}^2
    &= c_1^2\abs{i_{X(x)}\lambda}^2\\
    &\leq c_1^2 m\\
    &=m\,q^{-1}(c_1\lambda,c_1\lambda)\\
    &=m\,q^{-1}(c_1\lambda+c_2 n_x, c_1\lambda+c_2n_x).
\end{aligned}
\end{equation}
Where the first equality follows by $i_{X}n=0$. It follows that
\begin{equation}
\begin{aligned}
    \norm{\phi}^2_{H^1}
    &\leq \int_\cN \left((f+2)\abs{i_l d\phi}^2+(m+1)q^{-1}(d\phi,d\phi)+\abs{\phi}^2\right)\epsilon_\cN\\
    &\leq\max\{\max_{\cN}f+2,m+1,1\}\int_\cN\left(\abs{i_l d\phi}^2+q^{-1}(d\phi,d\phi)+\abs{\phi}^2\right)\epsilon_\cN.
\end{aligned}
\end{equation}

For the converse, we first seek a lower bound for \(i_X\lambda\). Since \(\lambda\in B_x\), there exists \(v\in T_x\cN\) such that \(\lambda=q(v)\). Hence,
\begin{equation}
    i_X\lambda=q(v,X)\geq -\sqrt{q(v,v)q(X,X)}\geq -\sqrt{f\,q(v,v)},
    \end{equation}
where the first inequality follows from the Cauchy-Schwarz inequality for the positive semidefinite form \(q\), and the second follows from the condition for positive semidefiniteness. As before, this inequality extends to every element of \(T_x^*\cN\), since \(i_X n_x=0\). Moreover, for every \(\nu=c_1\l +c_2 n_x\in T_x^*\cN\) with $\lambda=q(v)$, we have
\[q^{-1}(\nu,\nu)=c_1^2q^{-1}(\l,\l)=c_1^2q(v,v).\]
It follows that, for every \(\nu\in T_x^*\cN\),
\begin{equation}
    i_X\nu\geq-\sqrt{f\,q^{-1}(\nu,\nu)}.
\end{equation}
We can now estimate the derivative terms in the norm \eqref{eq:norm_Sob-scalar}:
\begin{equation}
\begin{aligned}
    f\abs{i_l d\phi}^2+q^{-1}(d\phi,d\phi)+2\,i_l d\phi\,i_Xd\phi&\geq f\abs{i_l d\phi}^2+q^{-1}(d\phi,d\phi)-2\sqrt{f}\,\abs{i_l d\phi}\sqrt{q^{-1}(d\phi,d\phi)}\\
    &=\left(\sqrt{f}\abs{i_l d\phi}-\sqrt{q^{-1}(d\phi,d\phi)}\right)^2\\
    &\geq(1-t)f\abs{i_ld\phi}^2+\left(1-\frac1t\right)q^{-1}(d\phi,d\phi),
\end{aligned}
\end{equation}
where the last inequality holds for every \(t>0\), by Young's inequality for products.\footnote{Specifically, for \(x,y\geq0\) and \(t>0\),\[2xy\leq tx^2+\frac{y^2}{t}.\]} Choosing
\[t=1+\frac{1}{f+1}\]
gives
\begin{equation}
\begin{aligned}
    \norm{\phi}^2_{H^1} &=\int_\cN\Big((f+1)\abs{i_l d\phi}^2+q^{-1}(d\phi,d\phi)+2\,i_l d\phi\,i_X d\phi+\abs{\phi}^2\Big)\epsilon_\cN\\
    &\geq\int_\cN\Big(\abs{i_l d\phi}^2+(1-t)f\abs{i_ld\phi}^2+\left(1-\tfrac1t\right)q^{-1}(d\phi,d\phi)+\abs{\phi}^2\Big)\epsilon_\cN\\
    &=\int_\cN\Big(\tfrac{1}{f+1}\abs{i_l d\phi}^2+\tfrac{1}{f+2}q^{-1}(d\phi,d\phi)+\abs{\phi}^2\Big)\epsilon_\cN\\
    &\geq\int_\cN\Big(\tfrac{1}{f+2}\abs{i_l d\phi}^2+\tfrac{1}{f+2}q^{-1}(d\phi,d\phi)+\abs{\phi}^2\Big)\epsilon_\cN\\
    &\geq \min\{1,\min_{\cN}\tfrac{1}{f+2}\}\int_\cN\Big(\abs{i_l d\phi}^2+q^{-1}(d\phi,d\phi)+\abs{\phi}^2\Big)\epsilon_\cN\ .
\end{aligned}
\end{equation}
This establishes that the norm induced by any such choice of generalized inverse is equivalent to the norm induced by the rigging inverse, although the two norms need not coincide. Consequently, they induce the same topology.

\medskip

To prove that the space of scalar fields vanishing at the boundary is closed, we follow the same strategy as in the Riemannian case: we define a trace operator and show that it is bounded. Recalling that \(\partial\cN\cong\bS\sqcup\bS\), we define
\begin{equation}
\begin{matrix}
T:&C^\infty(\cN)&\longrightarrow&
L^2(\bS,\epsilon_{S_0})
\oplus L^2(\bS,\epsilon_{S_1}),\\
&\phi&\longmapsto&
\left(
\phi|_{\bS_0},
\phi|_{\bS_1}
\right).
\end{matrix}
\end{equation}
Here $S_0$ and $S_1$ are the copies of the sphere in $\cN$ at $s=s_0$ and $s=s_1$, respectively.
The boundedness of this operator can be established by the same argument as in the Riemannian case. In particular,
\begin{equation}
\begin{split}
&\int_{\bS_0}\abs{\phi}^2\epsilon_{S_0}
+\int_{\bS_1}\abs{\phi}^2\epsilon_{S_1}\\
&\qquad=\int_\cN d\left(\abs{\phi}^2 i_l\epsilon_\cN\right)\\
&\qquad=\int_\cN\left(
2\phi\,d\phi\wedge i_l\epsilon_\cN
+\abs{\phi}^2d(i_l\epsilon_\cN)
\right)\\
&\qquad=\int_\cN\left(
2\phi\,i_ld\phi+\abs{\phi}^2\theta
\right)\epsilon_\cN\\
&\qquad\leq
\int_\cN\abs{\phi}^2\epsilon_\cN
+\int_\cN\abs{i_ld\phi}^2\epsilon_\cN
+\int_\cN\abs{\phi}^2\theta\,\epsilon_\cN\\
&\qquad\leq
\left(1+\max_\cN\abs{\theta}\right)
\int_\cN\left(
\abs{\phi}^2+\abs{i_ld\phi}^2
\right)\epsilon_\cN\\
&\qquad\leq
\left(1+\max_\cN\abs{\theta}\right)
\int_\cN\left(
\abs{\phi}^2+\abs{d\phi}^2+\abs{i_ld\phi}^2
\right)\epsilon_\cN,
\end{split}
\end{equation}
where \(\theta\) is the expansion defined in Eq.~\eqref{Thc} and is fixed by the background metric. This show that $T$ is bounded, and since the space of vanishing scalar field on the boundary is exactly its kernel. This space is closed.

\subsection{Converse case}
In analogy with the one-dimensional case, we would like to consider a norm such that the space of functions whose derivative vanishes at the boundary is not dense, while the space of functions vanishing at the boundary is it.

Let us start from the space of smooth functions $C^\infty(\cN)$. The hypersurface $\cN$ is foliated by a family of two-spheres, each equipped with an induced Riemannian metric $q_s$, parametrized smoothly by $s\in I$. For a fixed choice of inverse metric on each leaf, we can write
\begin{equation}\label{newphi}
\norm{\phi}^2=\oint_{\bS_0}
\abs{\dot{\phi}}^2\,\epsilon_{S}
+
\oint_{\bS_1}
\abs{\dot{\phi}}^2\,\epsilon_{S}
+
\int_\cN
\left(
\abs{\phi}^2+\abs{\Dd\phi}^2
\right)
\epsilon_\cN\,,
\end{equation}
and we recall that the area 2-form $\epsilon_{S}$ induced by the metric $q_{s}$ depends on $s$, $\abs{\Dd\phi}$ is the pointwise norm of the gradient of $\phi$ tangent to the two-spheres, computed with respect to the corresponding induced metric, and $\dot\phi$ denotes the derivative of the field along the $s$ direction.

As in the previous case, we aim to look to the kernel of the differential trace operator to find the properties of the space of functions whose derivative vanishes at the boundary. Then, let us define the differential trace operator
\begin{equation}
\begin{matrix}
DT:&C^\infty(\cN)&\longrightarrow&
L^2(\bS,\epsilon_{S_0})\oplus L^2(\bS,\epsilon_{S_1})\,;\\
&\phi&\longmapsto&
\left(
\dot\phi|_{\bS_0},
\dot\phi|_{\bS_1}
\right)\,.
\end{matrix}
\end{equation}
It is immediate that $DT$ is bounded with respect to the norm $\norm{\cdot}$. Consequently,
\begin{equation}
\left\{
\phi\in C^\infty(\cN)
\ \middle|\
\dot\phi|_{\bS_0}=\dot\phi|_{\bS_1}
=0
\right\}=\ker DT
\end{equation}
is closed.

\smallskip

The check of denseness of functions vanishing at the boundary is more involved. 
Let us first observe that, for every fixed $s\in I$, 
\[\phi(s)\coloneqq\phi|_{\bS_s}:\bS\longrightarrow\mathbb{R},\]
defines a smooth function on $\bS$, equipped with the Riemannian metric $q_s$, and hence an element of the associated Sobolev space $H^1(\bS,q_s)$. The collection of spaces $H^1(\bS,q_s)$ forms a measurable family of Hilbert spaces, and, with the smooth dependence of $q_s$ on $s$, may be regarded as a smooth Hilbert bundle over $I$. Hence, we may regard $\phi$ as a smooth section of this bundle, and thus as an element of
\begin{equation}
\int_I^\oplus H^1(\bS,q_s)\,ds.
\end{equation}
This bundle is trivial, and furthermore it is tensor product of $L^2$ functions on the interval with Sobolev functions on the sphere, as we now prove.

Since $\bS$ is compact, the Sobolev spaces associated with different smooth Riemannian metrics are identical as sets and have equivalent norms \cite{Hebey_1996}. We shall denote by $H^1(\bS)$ the Sobolev space associated with the standard metric on the sphere. Thus, as sets, the direct integral may be identified with \cite{Takesaki_1979}
\begin{equation}
L^2\bigl(I,H^1(\bS)\bigr).
\end{equation}
Let us make this identification more explicit. Since $H^1(\bS,q_s)$ and $H^1(\bS)$ are the same topological vector space, any smooth section of the Hilbert bundle may be regarded as an $H^1(\bS)$-valued smooth function. Moreover, the equivalence of the corresponding norms implies that, for every $s\in I$, there exist positive constants $\alpha_s$ and $\beta_s$ such that
\[\alpha_s \norm{f}_{H^1(\bS)}^2
\leq
\norm{f}_{H^1(\bS,q_s)}^2
\leq
\beta_s\norm{f}_{H^1(\bS)}^2.\]
Since $q_s$ depends smoothly on $s$ and $I$ is compact, these equivalences can be chosen uniformly in $s$. Namely, there exist constants $\alpha,\beta>0$ such that
\begin{equation}
    \alpha \norm{f}_{H^1(\bS)}^2
\leq
\norm{f}_{H^1(\bS,q_s)}^2
\leq
\beta\norm{f}_{H^1(\bS)}^2
\end{equation}
for every $s\in I$ and every $f\in H^1(\bS)$.

We recall that the norm of an element $\sigma$ in the direct integral is given by
\begin{equation}
\norm{\sigma}_{\int}^2
=
\int_I
\norm{\sigma(s)}^2_{H^1(\bS,q_s)}ds.
\end{equation}
Therefore,
\begin{equation}
\alpha\int_I
\norm{\sigma(s)}^2_{H^1(\bS)}ds
\leq
\norm{\sigma}_{\int}^2
\leq
\beta\int_I
\norm{\sigma(s)}^2_{H^1(\bS)}ds.
\end{equation}
The integral on the right-hand side is precisely the squared norm in
$L^2\bigl(I,H^1(\bS)\bigr)$. Hence, the two spaces are identical as sets and have equivalent norms. Moreover, since $H^1(\bS)$ is separable, one has the canonical isometric identification
\begin{equation}
L^2\bigl(I,H^1(\bS)\bigr)
\cong
L^2(I,ds)\otimes H^1(\bS).
\end{equation}

Having proved the tensor product structure, it suggests the following candidate for the completion:
\begin{equation}
\overline{C^\infty(\cN)}^{\scriptscriptstyle\norm{\cdot}}
\cong
L^2(\bS,\epsilon_{S_0})
\oplus
L^2(\bS,\epsilon_{S_1})\oplus\int_I^\oplus H^1(\bS,q_s)\,ds\ .
\end{equation}
At this point, the argument to establish denseness follows the same strategy as in the one-dimensional case. Namely, we consider the isometric embedding
\begin{equation}
\begin{matrix}
U:&C^\infty(\cN)&\longrightarrow&
L^2(\bS,\epsilon_{S_0}) \oplus L^2(\bS,\epsilon_{S_1})\oplus\int_I^\oplus H^1(\bS,q_s)\,ds\,;\\
&\phi&\longmapsto&
\left(
\phi,
\dot\phi|_{\bS_0},
\dot\phi|_{\bS_1}
\right).
\end{matrix}
\end{equation}

Writing the bulk integral as
\begin{equation}
\int_\cN
\left(
\abs{\phi}^2+\abs{\Dd\phi}^2
\right)
\epsilon_\cN=\int_I ds
\oint_{\bS}
\left(
\abs{\phi(s)}^2+\abs{\Dd\phi(s)}^2
\right)
\epsilon_{S_s}
=\int_Ids\norm{\sigma(s)}^2_{H^1(\bS,q_s)},
\end{equation}
we observe that the bulk contribution to $\norm{\phi}^2$ is precisely the norm of the first component of $U(\phi)$.

It remains to prove that the range of $U$ is dense. We proceed by two steps: first we prove that $C^\infty(N)$ is dense in $C^{\infty}(I,H^1(\bS))$ with respect to the norm $\norm{\cdot}$; and then that the completion of the latter is the candidate space.

Given any $\phi\in C^{\infty}(I,H^1(\bS))$, we may provide an approximation considering
\begin{equation}
    \phi_N (s)=\sum_{k=1}^N \braket{e_k,\phi(s)}_{H^1(\bS)}\,e_k
\end{equation}
where $e_k$ are a orthonormal basis of $H^1(\bS)$. Clearly such a sequence converges pointwise, i.e. for each fixed $s\in I$, to $\phi(s)$ in the Sobolev norm. Moreover,
\begin{equation}
    \norm{\phi_N(s)}_{H^1(\bS)}\leq \norm{\phi(s)}_{H^1(\bS)}\qquad \forall s\in I.
\end{equation}
Hence, by dominated convergence theorem $\phi_N\to \phi$ as elements of $L^2(I,H^1(\bS))$. It remains to show that it also converge in our norm $\norm{\cdot}$. Let us call
\begin{equation}
    f_k(s)=\braket{e_k,\phi(s)}_{H^1(\bS)}\,,
\end{equation}
and notice that it is a smooth function on $I$, moreover
\begin{equation}
    \dot f_k(s)=\braket{e_k,\dot\phi(s)}_{H^1(\bS)}\,,
\end{equation}
where the dot indicates the derivative with respect to $s$. Hence, 
\begin{equation}
    \dot\phi_N (s)=\sum_{k=1}^N \dot f_k(s)\,e_k=\sum_{k=1}^N \braket{e_k,\dot\phi(s)}_{H^1(\bS)}\,e_k
\end{equation}
and converges pointwise to $\dot\phi(s)$ in the Sobolev norm, and so, in particular, in $L^2$ norm
\begin{equation}
    \norm{\dot\phi(s)-\dot\phi_N(s)}_{L^2(\bS)}\leq\norm{\dot\phi(s)-\dot\phi_N(s)}_{H^1(\bS)}\longrightarrow 0\qquad \forall s\in I.
\end{equation}
In our norm we obtain
\begin{equation}
    \begin{split}
        \norm{\phi-\phi_N}^2&=\oint_{\bS}
\abs{\dot{\phi}(s_0)-\dot\phi_N(s_0)}^2\,\epsilon_{S_0}
+
\oint_{\bS}
\abs{\dot{\phi}(s_1)-\dot\phi_N(s_1)}^2\,\epsilon_{S_1}
+
\int_I ds\norm{\phi(s)-\phi_N(s)}_{H^1(\bS,q_s)}^2\\
&\leq \norm{\dot\phi(s_0)-\dot\phi_N(s_0)}_{L^2(\bS,\epsilon_{S_0})}^2+\norm{\dot\phi(s_1)-\dot\phi_N(s_1)}_{L^2(\bS,\epsilon_{S_1})}^2+\beta
\int_I ds\norm{\phi(s)-\phi_N(s)}_{H^1(\bS)}^2\\
&\leq \beta_0\norm{\dot\phi(s_0)-\dot\phi_N(s_0)}_{H^1(\bS)}^2+\beta_1\norm{\dot\phi(s_1)-\dot\phi_N(s_1)}_{H^1(\bS)}^2+\beta
\int_I ds\norm{\phi(s)-\phi_N(s)}_{H^1(\bS)}^2\\
&= \beta_0\norm{\dot\phi(s_0)-\dot\phi_N(s_0)}_{H^1(\bS)}^2+\beta_1\norm{\dot\phi(s_1)-\dot\phi_N(s_1)}_{H^1(\bS)}^2+\beta\norm{\phi-\phi_N}_{L^2(I,H^1(\bS))}^2\,.
    \end{split}
\end{equation}
It is easy to see that all the terms tend to zero by previous considerations. This proves that the space of finite linear combination of decomposable tensors is dense with respect our norm. Namely, the subspace given by algebraic tensor product 
\begin{equation}
    C^{\infty}(I)\otimes H^1(\bS)\hookrightarrow C^{\infty}(I,H^1(\bS))
\end{equation}
is dense with respect to the norm $\norm{\cdot}$. We complete the first stem of the proof by considering, for each $e_k\in H^1(\bS)$, a sequence of smooth functions on the sphere $e_{k,m}\in C^{\infty}(\bS)$ converging to $e_k$ in Sobolev norm, and the approximating sequence of functions
\begin{equation}
    \phi_{N,m}(s)=\sum_{k=1}^N f_k(s)e_{k,m}.
\end{equation}
Those are smooth functions on $\cN=I\times \bS$ and converge to $\phi_N$ in our norm
\begin{equation}
    \begin{split}
        \norm{\phi_N-\phi_{N,m}}^2&=\oint_{\bS}
\abs{\dot{\phi}_N(s_0)-\dot\phi_{N,m}(s_0)}^2\,\epsilon_{S_0}
+
\oint_{\bS}
\abs{\dot{\phi}_N(s_1)-\dot\phi_{N.m}(s_1)}^2\,\epsilon_{S_1}\\
&\quad+
\int_I ds\norm{\phi_N(s)-\phi_{N,m}(s)}_{H^1(\bS,q_s)}^2\\
&\leq \beta_0\norm{\dot\phi_N(s_0)-\dot\phi_{N,m}(s_0)}_{H^1(\bS)}^2+\beta_1\norm{\dot\phi_N(s_1)-\dot\phi_{N,m}(s_1)}_{H^1(\bS)}^2\\
&\quad+\beta\int_I ds\norm{\phi_N(s)-\phi_{N,m}(s)}_{H^1(\bS)}^2\\
&\leq\beta_0\sum_{k=1}^N\abs{\dot f_k(s_0)}^2\norm{e_k-e_{k,m}}^2_{H^1(\bS)}+\beta_1\sum_{k=1}^N\abs{\dot f_k(s_1)}^2\norm{e_k-e_{k,m}}^2_{H^1(\bS)}\\
&\quad+\beta\sum_{k=1}^N\int_I\abs{f_k(s)}^2\norm{e_k-e_{k,m}}^2_{H^1(\bS)}\\
&=\left(\beta_0\sum_{k=1}^N\abs{\dot f_k(s_0)}^2+\beta_1\sum_{k=1}^N\abs{\dot f_k(s_1)}^2+\beta\sum_{k=1}^N\int_I\abs{f_k(s)}^2\right)\norm{e_k-e_{k,m}}^2_{H^1(\bS)}
    \end{split}
\end{equation}
Where the factor in front of $\norm{e_k-e_{k,m}}^2_{H^1(\bS)}$ does not depend on $m$. Thus, the norm tends to zero. This chain of approximation shows that $C^{\infty}(\cN)$ is dense in $C^{\infty}(I,H^1(\bS))$ with respect our norm.

The second step is to prove that
\begin{equation}
\overline{C^{\infty}(I,H^1(\bS))}^{\scriptscriptstyle\norm{\cdot}}
\cong
L^2(\bS,\epsilon_{S_0})
\oplus
L^2(\bS,\epsilon_{S_1})\oplus\int_I^\oplus H^1(\bS,q_s)\,ds\ .
\end{equation}
Here, we want to use the fact that 
\[\int_I^\oplus H^1(\bS,q_s)\,ds\quad \text{and} \quad L^2(I,ds)\otimes H^1(\bS).\]
are equivalent. The map $U$ extends uniquely by continuity on $C^{\infty}(I,H^1(\bS))$. Consider a generic element
\[(\psi,\chi,\phi)
\in
L^2(\bS,\epsilon_{S_0}) \oplus L^2(\bS,\epsilon_{S_1})\oplus\int_I^\oplus H^1(\bS,q_s)\,ds\]
by equivalence and denseness, we may regard the function $\phi$ as a finite linear combination of decomposable tensors
\[\phi=\sum_{k=0}^n h_ke_k\,, \text{ for some } h_k\in L^2(I)\text{ and } e_k \text{ o.n.b. in } H^1(\bS)\,, \]
and,
\[\psi\in H^1(\bS)\subset L^2(\bS,\epsilon_{S_0}) \text{ and }\chi \in H^1(\bS)\subset L^2(\bS,\epsilon_{S_1}).\]
For each $h_k$ consider the sequence of smooth functions $h_{k,m}$ converging to $h_k$ in $L^2(I)$ and construct the sequence of functions
\begin{equation}
\phi_{k,m}(s)=
\begin{cases}
 h_{k,m}(s_0+\frac1m)\, e_k
+
\frac{1}{n}(s-s_0)\left[1-m(s-s_0)\right]\psi,
&
s_0\leq s<s_0+\frac1m,\\[2mm]
 h_{k,m}(s)\,e_k,
&
s_0+\frac1m\leq s\leq s_1-\frac1m,
\\[2mm]
h_{k,m}\left(s_1-\frac1m\right)\, e_k
+
\frac{1}{n}(s-s_1)\left[1+m(s-s_1)\right]\chi,
&
s_1-\frac1m<s\leq s_1,
\end{cases}
\end{equation}
These are $H^1(\bS)$-valued smooth functions on $I$ and so are they finite linear combinations
\[\phi_m=\sum_{k=1}^n\phi_{k,m}.\]
We want now prove that $U\phi_m$ converge to $(\psi,\chi,\phi)$. Notice that
\begin{equation}
    U\phi_m=(\psi,\chi,\phi_m)
\end{equation}
and so
\begin{align*}
    &\norm{(\psi,\chi,\phi)-U\phi_m}^2=\norm{\phi-\phi_m}^2_{\int}\\
    &=\int_I \norm{\phi(s)-\phi_m(s)}^2_{H^1(\bS,q_s)} ds\\
    &\leq\beta\int_I \norm{\phi(s)-\phi_m(s)}^2_{H^1(\bS)} ds\\
    &=\beta \int_{s_0}^{s_0+\frac{1}{m}}\norm{\phi(s)-\sum_{k=1}^nh_{k,m}(s_0+\tfrac1m)\, e_k-(s-s_0)\left[1-m(s-s_0)\right]\psi}^2_{H^1(\bS)} ds\\
    &\qquad +\beta\int_{s_0+\frac{1}{m}}^{s_1-\frac{1}{m}}\norm{\phi(s)-\sum_{k=1}^nh_{k,m}(s)\, e_k}^2_{H^1(\bS)} ds\\
    &\qquad +\beta \int_{s_1-\frac{1}{m}}^{s_1}\norm{\phi(s)-\sum_{k=1}^n h_{k,m}\left(s_1-\tfrac1m\right)\, e_k-(s-s_1)\left[1+m(s-s_1)\right]\chi}^2_{H^1(\bS)} ds\\
    &\leq2\beta \int_{s_0}^{s_0+\frac{1}{m}}\left(\sum_{k=1}^n\norm{h_k(s)e_k-h_{k,m}(s_0+\tfrac1m)\, e_k}^2_{H^1(\bS)}+\norm{(s-s_0)\left[1-m(s-s_0)\right]\psi}^2_{H^1(\bS)}\right)ds\\
    &\qquad +2\beta\int_{s_0+\frac{1}{m}}^{s_1-\frac{1}{m}}\sum_{k=1}^n\norm{h_k(s)e_k-h_{k,m}(s)\, e_k}^2_{H^1(\bS)}ds\\
    &\qquad +2\beta \int_{s_1-\frac{1}{m}}^{s_1}\left(\sum_{k=1}^n\norm{h_k(s)e_k- h_{k,m}\left(s_1-\tfrac1m\right)\, e_k}^2_{H^1(\bS)}+\norm{(s-s_1)\left[1+m(s-s_1)\right]\chi}^2_{H^1(\bS)}\right)ds\\
    &=2\beta \sum_{k=1}^n\norm{e_k}^2_{H^1(\bS)}\int_{s_0}^{s_0+\frac{1}{m}}\abs{h_k(s)-h_{k,m}(s_0+\tfrac1m)}^2\,ds\\
    &\qquad +2\beta\sum_{k=1}^n\norm{e_k}^2_{H^1(\bS)}\int_{s_0+\frac{1}{m}}^{s_1-\frac{1}{m}}\abs{h_k(s)-h_{k,m}(s)}^2ds\\
    &\qquad +2\beta \sum_{k=1}^n\norm{e_k}^2_{H^1(\bS)}\int_{s_1-\frac{1}{m}}^{s_1}\abs{h_k(s)- h_{k,m}\left(s_1-\tfrac1m\right)}^2 ds\\
    &\qquad+2\beta\norm{\psi}^2_{H^1(\bS)}\int_{s_0}^{s_0+\frac{1}{m}}\abs{(s-s_0)\left[1-m(s-s_0)\right]}^2ds\\
    &\qquad+2\beta\norm{\chi}^2_{H^1(\bS)}\int_{s_1-\frac{1}{m}}^{s_1}\abs{(s-s_1)\left[1+m(s-s_1)\right]}^2 ds \\
    &\leq4\beta \sum_{k=1}^n\norm{e_k}^2_{H^1(\bS)}\norm{h_k-h_{k,m}}^2_{L^2(\bS)}\\
    &\qquad+4\beta \sum_{k=1}^n\norm{e_k}^2_{H^1(\bS)}\left(\int_{s_0}^{s_0+\frac{1}{m}}\abs{h_{k,m}(s)-h_{k,m}(s_0+\tfrac1m)}^2\,ds+\int_{s_1-\frac{1}{m}}^{s_1}\abs{h_{k,m}(s)- h_{k,m}(s_1-\tfrac1m)}^2 ds\right)\\
    &\qquad+\frac{2\beta\norm{\psi}^2_{H^1(\bS)}}{16m^3}+\frac{2\beta\norm{\chi}^2_{H^1(\bS)}}{16m^3}\ .
\end{align*}
These terms vanish similarly to the one-dimensional case. It follows that the range of $U$ is dense, which therefore gives the completion of $C^\infty(\cN)$ with respect to the norm $\norm{\cdot}$. Notice that the boundary-layer construction only requires Sobolev regularity of the boundary data at the intermediate stage; arbitrary $L^2$ boundary data are then obtained by density.

\subsection{Mixed case}

Let us now consider the 1-parameter family \eqref{Ixiphic}. We restrict attention to the choices properly associated with polarizations, see footnote \ref{FootPol}.
In these cases, one can show that replacing the boundary values in \eqref{newphi} with $|\cos\th\,\dot\phi - \sin\th\,\phi|^2$,  makes the vanishing of this linear combination a closed subset, and the vanishing of $\cos\th\,\phi + \sin\th\,\dot\phi$ dense. We therefore recover the Sobolev and converse topologies from $\th=\pi/2$ and $\th=0$ respectively.\footnote{The norm obtained for $\th=\pi/2$ differs from \eqref{eq:norm_Sob-scalar} by an additional Euclidean norm $|\phi|^2$ term, which does not affect the topology.}

\section{Geometries on null hypersurface}
This case presents two fundamental differences with respect to the scalar field of the previous example, and to the null infinity framework of \cite{Ashtekar:1981bq,Ashtekar:2024stm} as well. 

The first is that the space of dynamical fields, the metrics, is not a linear space, hence we cannot define a norm intrinsically on it.
Let \(\cN=I\times\bS\). The space of Riemannian metrics on \(\cN\) is a convex cone, i.e., a space closed under linear combinations with positive coefficients, in the space of symmetric tensors,
\begin{equation}
\mathrm{Met}(\cN)=\left\{ q \in \Gamma(\mathrm{S}^2 T^*\cN) \ \middle|\ q>0 \right\}.
\end{equation}
The degenerate metrics lie on the boundary of this convex cone and form a cone of their own, i.e., a space closed under multiplication by a positive factor (for details cf. Appendix~\ref{App}). The space we are interested in is the space of degenerate metrics whose kernel is exactly given by the direction \(l\), corresponding to the geometries of the null hypersurface \(\cN\):
\begin{equation}
\mathrm{nGeom}(\cN)\coloneqq
\left\{q \in \Gamma(\mathrm{S}^2 T^*\cN)\ \middle|\ q\geq0,\, \ker q = \mathrm{Span}(l)\right\}.
\end{equation}
Again, this space is a convex cone. Recall that the metric \(q\)  always descends to a positive-definite metric on the quotient bundle $T\cN/\mathrm{Span}(l)$. Since the latter is isomorphic to $TS$, the space of null geometries can be identified with the space of smooth maps from \(I\) to the space of Riemannian metrics on \(\bS\):
\begin{equation}
\mathrm{nGeom}(\cN)
\cong
C^\infty\bigl(I,\mathrm{Met}(\bS)\bigr).
\end{equation}
This is the space of dynamical fields of interest. It is not a linear space, and in fact none of the above spaces of metrics are linear. They are however all subspaces of the linear space $\Gamma(\mathrm{S}^2 T^*\cN)$. Therefore, even if it is not possible to define a norm on them intrinsically, we can equip them with a notion of distance induced by a norm on the ambient linear space, and discuss convergence with respect to this distance.

The second difference is that since the geometry is dynamical, we must introduce an \emph{auxiliary} metric in order to define the volume form and the point-wise norm, that we will use to construct norms on the (ambient) space of geometries.\footnote{This difference is more nuanced: in fact, even in the scalar field example and in the construction at $\scri$, one could have used arbitrary background metrics to induce the norms. It is just natural to use the one already available from the geometry of the hypersurface.} It is natural to choose the background metric also degenerate, in order for it to respect the split \eqref{NIS} and allows use to distinguish boundary from bulk contributions. As in the scalar field case, any choice of such background metric will gives rise to an equivalent topology. We choose for simplicity the pull-back on $\cN$ of the standard unit-round sphere Riemannian metric \(\hat{q}\) on \(\bS\), and we will denote $\hat\Dd$ it covariant derivative.

\subsection{Sobolev case}

In the standard case, one seeks a norm for which geometries that vanish at the boundary are not dense, while no control is imposed on the derivatives of the geometries at the boundary. This is precisely the type of behavior provided by a Sobolev norm. It is therefore natural to define a norm on the linear space $C^\infty(I,\Gamma(\mathrm{S}^2 T^*\bS))$ as 
\begin{equation}
\label{eq:norm-Sob}
\norm{t}^2_{H^1}=\int_I ds \oint_{\bS}\left(\abs{t}^2+\abs{\hat{\Dd}t}^2+\abs{\dot{t}}^2\right)\epsilon_{\hat{q}}\,,
\end{equation}
where the absolute value denotes the pointwise norm induced by \(\hat q\) on the appropriate vector bundle: \(\mathrm{S}^2T^*\bS\) for \(t\) and \(\dot t\), and \(T^*\bS\otimes\mathrm{S}^2T^*\bS\) for \(\hat\Dd t\). In coordinates, this norm reads
\[\norm{t}^2_{H^1}=\int_\cN\left(\hat{q}^{AC}\hat{q}^{BD}t_{AB}t_{CD}+\hat{q}^{AD}\hat{q}^{BE}\hat{q}^{CF}{\hat{\Dd}}_At_{BC}{\hat{\Dd}}_Dt_{EF}+\hat{q}^{AC}\hat{q}^{BD}\dot t_{AB}\dot t_{CD}\right)\sqrt{\hat{q}}\,d^3x\,.\]
The distance between two geometries is then defined by
 \[d(q,q')=\norm{q-q'}_{H^1}\]
Although \(\mathrm{nGeom}(\cN)\) is not itself a vector space, this distance is well defined because it is induced by the norm on the ambient linear space of symmetric tensors. It therefore equips \(\mathrm{nGeom}(\cN)\) with the topology induced by the corresponding \(H^1\)-type norm.

\smallskip

Notice that the norm \eqref{eq:norm-Sob} can equivalently be obtained by equipping the space $
\Gamma(\mathrm{S}^2T^*\cN)$ with a Sobolev norm associated with an auxiliary Riemannian metric on \(\cN\). More precisely, one can choose a Riemannian metric on \(\cN\) whose restriction to the transverse directions coincides with \(\hat q\) and for which \(l\) is a unit vector field. The resulting Sobolev norm is equivalent to \eqref{eq:norm-Sob}. Hence, there is a set-theoretic and topological embedding
\begin{equation}
\mathrm{nGeom}(N)\hookrightarrow H^1(\cN,\mathrm{S}^2T^*\cN).
\end{equation}
Here and below, by an embedding we simply mean that the topology on \(\mathrm{nGeom}(\cN)\) is the one induced by the corresponding norm on the ambient Sobolev space.

In this setting, it is straightforward to prove that the space of geometries vanishing at the boundary is closed. Let us consider the trace operator
\begin{equation}
\begin{matrix}
T:&C^\infty(I,\Gamma(\mathrm{S}^2 T^*\bS))&\longrightarrow& L^2(\bS,\mathrm{S}^2T^*\bS)\oplus L^2(\bS,\mathrm{S}^2T^*\bS)\,;\\
&t&\longmapsto&\left(t|_{\bS_0},t|_{\bS_1}\right)\,.
\end{matrix}
\end{equation}
The \(L^2\) spaces in this expression are defined using the auxiliary metric \(\hat q\). Since \(H^1\) admits a continuous trace on the boundary, one expects \(T\) to be continuous. We now verify this explicitly in the present setting.

Let us define two smooth functions \(\rho_0,\rho_1\) on the interval \(I\) such that
\[0\leq\rho_0\leq1,\qquad\rho_0(s_0)=1,\qquad \rho_0(s_1)=0,\qquad\dot\rho_0\leq0,\]
and such that all its derivatives in $s_0$ and $s_1$ are null.\footnote{A smooth function with the required properties can be constructed using the standard smooth cutoff function \cite{Lee_2003},
\[\rho(x)=\frac{e^{-\frac{1}{1-x}}}{e^{-\frac{1}{1-x}}+e^{-\frac{1}{x}}} \quad \text{for } x\in [0,1].\]}
Similarly, we choose \(\rho_1\) such that
\[0\leq\rho_1\leq1,\qquad\rho_1(s_0)=0,\qquad \rho_1(s_1)=1,\qquad\dot\rho_1\geq0,\]
Considering \(t\) as an element of $C^\infty(I,\Gamma(\mathrm{S}^2 T^*\bS)),$ we obtain
\begin{equation}
    \begin{split}
        &\oint_{\bS}\abs{t(s_0)}^2\epsilon_{\hat{q}}+\oint_{\bS}\abs{t(s_1)}^2\epsilon_{\hat{q}}=\oint_{\bS}\rho_0(s_0)\abs{t(s_0)}^2\epsilon_{\hat{q}}+\oint_{\bS}\rho_1(s_1)\abs{t(s_1)}^2\epsilon_{\hat{q}}\\
        &=\int^{s_0}_{s_1}\left(\frac{d}{ds}\oint_{\bS}\rho_0(s)\abs{t(s)}^2\epsilon_{\hat{q}}\right) ds+\int^{s_1}_{s_0}\frac{d}{ds}\left(\oint_{\bS}\rho_1(s)\abs{t(s)}^2\epsilon_{\hat{q}}\right)ds\\
        &=\int^{s_0}_{s_1}ds\oint_{\bS}\left(\dot\rho_0(s)\abs{t(s)}^2+\rho_0(s)\frac{d}{ds}\abs{t(s)}^2\right)\epsilon_{\hat{q}}+\int^{s_1}_{s_0}ds\oint_{\bS}\left(\dot\rho_1(s)\abs{t(s)}^2+\rho_1(s)\frac{d}{ds}\abs{t(s)}^2\right)\epsilon_{\hat{q}}\\
        &=\int^{s_1}_{s_0}ds\oint_{\bS}\left((\abs{\dot{\rho}_0(s)}+\dot\rho_1(s))\abs{t(s)}^2+(\rho_1(s)-\rho_0(s))\frac{d}{ds}\abs{t(s)}^2\right)\epsilon_{\hat{q}}
    \end{split}
\end{equation}
In the last equality, we used \(\dot\rho_0\leq0\), so that $\dot\rho_0=-\abs{\dot\rho_0}$. Taking absolute values in the second term and using the boundedness of the cutoff functions gives
\begin{equation}
    \begin{split}
        &\oint_{\bS}\abs{t(s_0)}^2\epsilon_{\hat{q}}+\oint_{\bS}\abs{t(s_1)}^2\epsilon_{\hat{q}}\\
        &\leq\int^{s_1}_{s_0}ds\oint_{\bS}\left((\abs{\dot{\rho}_0(s)}+\dot\rho_1(s))\abs{t(s)}^2+\abs{\rho_1(s)-\rho_0(s)}\left|\frac{d}{ds}\abs{t(s)}^2\right|\right)\epsilon_{\hat{q}}\\
        &\leq \max_{s\in I}(\abs{\dot{\rho}_0(s)}+\dot\rho_1(s))\int^{s_1}_{s_0}ds\oint_{\bS}\abs{t(s)}^2\epsilon_{\hat{q}}+\max_{s\in I}\abs{\rho_1(s)-\rho_0(s)}\int^{s_1}_{s_0}ds\oint_{\bS}\left|\frac{d}{ds}\abs{t(s)}^2\right|\epsilon_{\hat{q}}
    \end{split}
\end{equation}
Since the pointwise norm is induced by the fixed metric \(\hat q\), we have
\[\left|\frac{d}{ds}\abs{t(s)}^2\right|\leq2\abs{t(s)}\abs{\dot t(s)}\leq\abs{t(s)}^2+\abs{\dot t(s)}^2,\]
Consequently,
\begin{equation}
    \begin{split}
        &\oint_{\bS}\abs{t(s_0)}^2\epsilon_{\hat{q}}+\oint_{\bS}\abs{t(s_1)}^2\epsilon_{\hat{q}}\\
        &\leq \max_{s\in I}(\abs{\dot{\rho}_0(s)}+\dot\rho_1(s))\int^{s_1}_{s_0}ds\oint_{\bS}\abs{t(s)}^2\epsilon_{\hat{q}}+\max_{s\in I}\abs{\rho_1(s)-\rho_0(s)}\int^{s_1}_{s_0}ds\oint_{\bS}\abs{t(s)}^2\epsilon_{\hat{q}}\\
        &\qquad+\max_{s\in I}\abs{\rho_1(s)-\rho_0(s)}\int^{s_1}_{s_0}ds\oint_{\bS}\abs{\dot{t}(s)}^2\epsilon_{\hat{q}}\\
        &\leq \left(\max_{s\in I}(\abs{\dot{\rho}_0(s)}+\dot\rho_1(s))+\max_{s\in I}\abs{\rho_1(s)-\rho_0(s)}\right)\int^{s_1}_{s_0}ds\oint_{\bS}\left(\abs{t(s)}^2+\abs{\dot{t}(s)}^2\right)\epsilon_{\hat{q}}\\
        &\leq \left(\max_{s\in I}(\abs{\dot{\rho}_0(s)}+\dot\rho_1(s))+\max_{s\in I}\abs{\rho_1(s)-\rho_0(s)}\right)\int^{s_1}_{s_0}ds\oint_{\bS}\left(\abs{t(s)}^2+\abs{\hat{\Dd}t(s)}^2+\abs{\dot{t}(s)}^2\right)\epsilon_{\hat{q}}
    \end{split}
\end{equation}
This provides the desired estimate. In particular, there exists a constant \(C>0\), depending only on the choice of the cutoff functions, such that
\begin{equation}
\label{eq:trace-estimate-nGeom}
\|Tt\|_{L^2}^2\leq C\|t\|^2_{H^1}.\end{equation}
Hence, \(T\) is continuous with respect to the topology induced by the norm \eqref{eq:norm-Sob}. It follows that
\[\ker T=\left\{t\in C^\infty(I,\Gamma(\mathrm{S}^2 T^*\bS))\ \middle|\ t(s_0)=t(s_1)=0\right\}\]
is closed in \( C^\infty(I,\Gamma(\mathrm{S}^2 T^*\bS))\). Notice that, by definition, 
\begin{equation}
    Tq\neq 0\quad \text{for any } q\in \mathrm{nGeom}(\cN)
\end{equation}
since $q$ at the boundary must be a metric on the sphere. However, this by itself is not sufficient to exclude the possibility that, in the completion of the space with respect to the induced distance, the subset of null geometries that are fully degenerate at the boundary becomes dense. For this, we need the kernel of \(T\) to be closed. Indeed, suppose that \(q_n\) is a sequence of null geometries that are fully degenerate at the boundary, so that, in particular, \(q_n\in C^\infty(I,\Gamma(\mathrm{S}^2 T^*\bS))\), and that \(q_n\) converges to \(q\) with respect to \(\norm{\cdot}_{H^1}\). Since \(T\) is continuous, we have
\[Tq=\lim_{n\to\infty}Tq_n=0.\]
Thus, \(q\) also belongs to the kernel of \(T\), showing that the set of null geometries that are fully degenerate at the boundary is closed with respect to the induced distance. Thus, a sequence of geometries with vanishing boundary metric cannot approximate, with respect to the norm \eqref{eq:norm-Sob}, a geometry with non-vanishing boundary metric. Thus, in the standard Ashtekar case, the \(H^1\)-type norm retains the metric data at the boundary even though it does not control the boundary values of the derivatives.

\subsection{Converse case}
For the converse case, we take what we learn in the previous Section. We define a norm on $C^\infty(I,\Gamma(\mathrm{S}^2 T^*\bS))$ as 
\begin{equation}
\label{eq:norm-bdy}
\norm{t}^2=\oint_{\bS}\abs{\dot{t}(s_0)}^2\epsilon_{\hat{q}}+\oint_{\bS}\abs{\dot{t}(s_1)}^2\epsilon_{\hat{q}}+\int_I ds \oint_{\bS}\left(\abs{t(s)}^2+\abs{\hat{\Dd}t(s)}^2\right)\epsilon_{\hat{q}}\,.
\end{equation}
In this case the differential trace is continuous 
\begin{equation}
\begin{matrix}
DT:&\mathrm{nGeom}(\cN)&\longrightarrow& L^2(\bS,\mathrm{S}^2T^*\bS)\oplus L^2(\bS,\mathrm{S}^2T^*\bS)\,;\\
&q&\longmapsto&\left(\dot{q}|_{\bS_0},\dot{q}|_{\bS_1}\right)\,.
\end{matrix}
\end{equation}
Indeed,
\begin{equation}
    \norm{DT(q)}^2_{L^2}=\oint_{\bS}\abs{\dot{q}(s_0)}^2\epsilon_{\hat{q}}+\oint_{\bS}\abs{\dot{q}(s_1)}^2\epsilon_{\hat{q}}\leq \norm{q}^2
\end{equation}
Providing the usual close condition on metrics with vanishing derivatives. 

\smallskip

Unfortunately, the vanishing-at-the-boundary metrics are not dense. The requirement that \(h(s)\) be a metric on \(\bS\) at every point of the interior of \(I\) imposes conditions on its derivatives at the boundary, and these derivatives are explicitly controlled by the norm. We first formulate the problem in a larger space. This is necessary because, as observed above, the space of vanishing-at-the-boundary metrics is not a subset of \(\mathrm{nGeom}(\cN)\), although it may be contained in its completion.

We therefore split the argument into two steps. First, we show that the space of possibly degenerate geometries
\[C^\infty\bigl(I,\mathrm{Deg}(\bS)\bigr),\]
where
\begin{equation}
\mathrm{Deg}(\bS)=\left\{q\in\Gamma(\mathrm{S}^2T^*\bS)\,\middle|\,q\geq 0\right\},
\end{equation}
is contained in the completion
\begin{equation}
X=\overline{C^\infty\bigl(I,\mathrm{Met}(\bS)\bigr)}^{\scriptscriptstyle d}
\end{equation}
with respect to the distance
\begin{equation}
\label{eq:dist}
d(q,q')=\norm{q-q'}.
\end{equation}
We then show that, even in this enlarged space \(X\), there is an obstruction in approximating every $q\in C^\infty(I,\mathrm{Met}(\bS))$ by a sequence of vanishing-at-the-boundary metrics. Let us first note that
\[C^\infty\bigl(I,\mathrm{Deg}(\bS)\bigr)\]
is a convex cone containing
\[C^\infty\bigl(I,\mathrm{Met}(\bS)\bigr).\]
Moreover, it is closed and has empty interior with respect to the topology induced by the norm under consideration (see Appendix~\ref{App}). Notice also that elements of \(\mathrm{Deg}(\bS)\) need not have constant rank. In particular, this space contains elements \(h\) satisfying
\begin{equation}
\label{eq:null-bdy-met}
h(s)\in\mathrm{Met}(\bS)\,,\ \forall s\in(s_0,s_1),\qquad h(s_0)=h(s_1)=0.
\end{equation}
For the first step, we show that every
\[t\in C^\infty\bigl(I,\mathrm{Deg}(\bS)\bigr)\]
can be approximated by elements of
\[C^\infty\bigl(I,\mathrm{Met}(\bS)\bigr).\]
Indeed, let $q\in C^\infty(I,\mathrm{Met}(\bS))$ be arbitrary and consider the sequence
\begin{equation}
q_n(s)=\frac{1}{n}q(s)+t(s).
\end{equation}
Since \(q(s)\) is positive definite and \(t(s)\) is positive semidefinite, \(q_n(s)\) is positive definite for every \(s\in I\). Hence
\[q_n\in C^\infty\bigl(I,\mathrm{Met}(\bS)\bigr)\]
for every \(n\). Moreover,
\begin{equation}
d(q_n,t)^2=\norm{q_n-t}^2=\frac{1}{n^2}\norm{q}^2,
\end{equation}
which converges to zero as \(n\to\infty\), since \(\norm{q}\) is finite and independent of \(n\). Thus,
\[C^\infty\bigl(I,\mathrm{Met}(\bS)\bigr)\]
is dense in
\[C^\infty\bigl(I,\mathrm{Deg}(\bS)\bigr)\]
with respect to the distance \eqref{eq:dist}. Consequently,
\begin{equation}
C^\infty\bigl(I,\mathrm{Met}(\bS)\bigr)\subset C^\infty\bigl(I,\mathrm{Deg}(\bS)\bigr)\subset X,
\end{equation}
with
\begin{equation}
X=\overline{C^\infty\bigl(I,\mathrm{Deg}(\bS)\bigr)}^{\scriptscriptstyle d}=\overline{C^\infty\bigl(I,\mathrm{Met}(\bS)\bigr)}^{\scriptscriptstyle d}.
\end{equation}

We now show that a sequence \(h_n\) satisfying \eqref{eq:null-bdy-met} cannot approximate every $q\in C^\infty\bigl(I,\mathrm{Met}(\bS)\bigr).$ The condition \eqref{eq:null-bdy-met} implies
\begin{equation}
\dot h(s_0)\geq0,\qquad\dot h(s_1)\leq0.
\end{equation}
Indeed, for every \(x\in\bS\) and \(v\in T_x\bS\),
\begin{subequations}
\begin{equation}
\label{eq:neg-dot}
\dot h(s_0)_x(v,v)=\lim_{s\to s_0^+}\frac{h(s)_x(v,v)-h(s_0)_x(v,v)}{s-s_0}=\lim_{s\to s_0^+}\frac{h(s)_x(v,v)}{s-s_0}\geq0\,,
\end{equation}
\begin{equation}
\label{eq:pos-dot}
\dot h(s_1)_x(v,v)
=
\lim_{s\to s_1^-}
\frac{h(s_1)_x(v,v)-h(s)_x(v,v)}{s_1-s}
=
-\lim_{s\to s_1^-}
\frac{h(s)_x(v,v)}{s_1-s}
\leq0\,,
\end{equation}
\end{subequations}
for all $x\in\bS$ and $v\in T_x\bS$. Hence, for every \(n\),
\begin{equation}
\dot h_n(s_0)\in\mathrm{Deg}(\bS),\qquad-\dot h_n(s_1)\in\mathrm{Deg}(\bS).
\end{equation}
Suppose now that a sequence \(h_n\) of vanishing-at-the-boundary metrics converges to a metric
\[q\in C^\infty\bigl(I,\mathrm{Met}(\bS)\bigr)\]
with respect to the norm \eqref{eq:norm-bdy}. Since the norm contains the \(L^2\)-norms of the boundary derivatives, we have
\[\norm{\dot q(s_0)-\dot h_n(s_0)}_{L^2(\bS,\mathrm{S}^2T^*\bS)}\xrightarrow[n\to+\infty]{}0\,,\]
and
\[\norm{\dot q(s_1)-\dot h_n(s_1)}_{L^2(\bS,\mathrm{S}^2T^*\bS)}
\xrightarrow[n\to+\infty]{}0\,.\]
Since \(\mathrm{Deg}(\bS)\) is closed in the \(L^2\) topology (see Appendix~\ref{App}), it follows that
\[\dot q(s_0)\in\mathrm{Deg}(\bS)\,,\qquad-\dot q(s_1)\in\mathrm{Deg}(\bS)\,.\]
Thus, any metric \(q\) that can be obtained as such a limit must satisfy these additional boundary conditions. Since a generic element of
\[C^\infty\bigl(I,\mathrm{Met}(\bS)\bigr)\]
does not satisfy them, the vanishing-at-the-boundary metrics cannot be dense.

One might try to weaken the boundary condition by requiring only
\begin{equation}
\label{eq:null-bdy-met-bis}
h(s)\in\mathrm{Met}(\bS),\,\forall s\in(s_0,s_1),\qquad\operatorname{rk}h(s_0)\leq1,\quad\operatorname{rk}h(s_1)\leq1.
\end{equation}
However, this is still not sufficient. If \(v\in\ker h(s_0)_x\), then the argument leading to \eqref{eq:neg-dot} gives
\[\dot h(s_0)_x(v,v)\geq0.\]
Similarly, if \(v\in\ker h(s_1)_x\), then \eqref{eq:pos-dot} gives
\[\dot h(s_1)_x(v,v)\leq0.\]
Since a tensor of rank at most one on the two-dimensional space \(T_x\bS\) necessarily has a nontrivial kernel, this means that \(\dot h(s_0)_x\) has at least one nonnegative direction, while \(\dot h(s_1)_x\) has at least one nonpositive direction. Equivalently,
\[\dot h(s_0)_x
\quad\text{is not negative definite},
\qquad
\dot h(s_1)_x
\quad\text{is not positive definite}.\]
Therefore, if a sequence \(h_n\) satisfying \eqref{eq:null-bdy-met-bis} converges to \(q\) in the norm \eqref{eq:norm-bdy}, the same closedness argument implies
\[\dot q(s_0)_x
\quad\text{is not negative definite},
\qquad
\dot q(s_1)_x
\quad\text{is not positive definite},\]
for every \(x\in\bS\). Hence, even this weaker boundary condition imposes nontrivial restrictions on the possible limits.

In conclusion, enlarging the space from nondegenerate geometries to the cone
\[C^\infty\bigl(I,\mathrm{Deg}(\bS)\bigr)\]
does allow us to formulate the approximation problem in a space containing the relevant boundary-degenerate configurations. However, it does not restore density of the vanishing-at-the-boundary metrics. The obstruction is not merely a consequence of working in a space that is too small or of the formal properties of the completion: it arises from the pointwise positivity condition itself. Because the metric is required to remain positive definite in the interior of \(I\), its boundary behavior constrains its first derivative, and these constraints are detected by the boundary terms in the norm. Thus, the cone structure and the positivity requirement provide a genuine obstruction to the desired density result.

\section{Conclusions}

We have studied topologies in the radiative phase space, associated with a physical null hypersurface, of a scalar field and of the gravitational field. For both fields, we have provided new details on how the topology induced by the Sobolev norm selects a unique Hamiltonian generating diffeomorphisms, in spite of the `non-integrability' problem. This Hamiltonian can be understood as a Noether charge in the momentum polarization. For the scalar field, we have also shown that there exists an alternative choice of topology, induced by a new norm, that selects instead an Hamiltonian corresponding to the configuration polarization. In fact, this new norm can be readily generalized to a family that permits to include mixed Robin-type polarizations. However, extending this norm to the gravitational field does not work. Two challenges - the metric being dynamical and the space of metrics being non-linear - can be successfully addressed. It is the (semi)positivity of the metric that prevents the norm from providing denseness for the desired subset. Therefore in so far as the gravitational case is concerned, the CFP dynamical entropy appears at the moment to be preferred from such topological considerations.

We conclude with two open questions. First, we stress that we do not have a no-go theorem for the alternative polarizations in the gravitational case. Simply, the natural norm that works for functions and scalar fields, fails to apply, and while we have identified why, we could not find a viable alternative. 
Second, at least for the scalar, it would be interesting to study the continuity of the symplectic 2-form and Hamiltonians in the new topology.

\subsection*{Acknowledgements}

Simone thanks the organisers of the Jurek Memorial conference for the remembrance, and a stimulating atmosphere to continue on the research he inspired.  M.B. acknowledges support by grants ID\# 63132 and ID\# 63683 from the John Templeton Foundation, managed by the Center for SpaceTime and the Quantum.

\appendix
\section{Cone of metrics}
\label{App}
Let us consider the space of Riemannian metrics on a smooth manifold \(M\),
\begin{equation}
\mathrm{Met}(M)=\left\{g\in \Gamma(\mathrm{S}^2T^*M)\ \middle|\ g>0\right\},
\end{equation}
where \(g>0\) means that
\[g_x(v,v)>0 \qquad \forall x\in M,\quad \forall v\in T_xM\setminus\{0\}.\]
This space is equipped with the topology inherited from the compact-open topology on the space of smooth sections \(\Gamma(\mathrm{S}^2T^*M)\) of the symmetric cotangent bundle \(\mathrm{S}^2T^*M\). For the purposes of this work, we assume that \(M\) is compact. Then the compact-open topology is the topology of uniform convergence and it is metrizable by the distance
\begin{equation}
d(g,h)=\max_{x\in M}\max_{\substack{u\in T_xM \\ \abs{u}=1}}
\abs{g_x(u,u)-h_x(u,u)},
\end{equation}
where \(\abs{u}\) denotes the norm induced on \(T_xM\) by an auxiliary metric. By compactness of \(M\), the topology does not depend on the choice of auxiliary metric. With respect to this topology, the space that includes degenerate metrics,
\begin{equation}
\mathrm{Deg}(M)=\left\{g\in \Gamma(\mathrm{S}^2T^*M)\ \middle|\ g\geq0\right\},
\end{equation}
is a convex cone whose interior is \(\mathrm{Met}(M)\). The proof is standard, but for the sake of completeness, we report it here.

\begin{proof}
Let \(h\in \mathrm{Deg}(M)\setminus \mathrm{Met}(M)\). Then there exists at least one point \(x_*\in M\) and a tangent vector \(v\neq 0\in T_{x_*}M\) such that
\[h_{x_*}(v,v)=0.\]
Considering any smooth extension \(\tilde{v}\) of \(v\) to the whole of \(M\), we can construct a symmetric tensor, labeled by a positive real number \(\delta\),
\begin{equation}
t^\delta=h-\delta\,\tilde{v}^{\flat}\otimes\tilde{v}^{\flat},
\end{equation}
where the musical isomorphism is defined with respect to the auxiliary metric that induces the norm. We can see that, by varying \(\delta\), such tensors can be made arbitrarily close to \(h\); more precisely,
\[\forall\varepsilon>0\ \exists\delta>0\text{ such that }d(h,t^\delta)<\varepsilon.\]
Indeed,
\begin{equation}
d(h,t^\delta)=\delta\max_{x\in M}\max_{\substack{u\in T_xM\\ \abs{u}=1}}\abs{\tilde{v}_{x}^{\flat}(u)}^2\leq \delta\max_{x\in M}\abs{\tilde{v}_x}^2,
\end{equation}
because $\abs{\tilde{v}_x^{\flat}(u)}^2\leq \abs{\tilde{v}_x}^2\abs{u}^2.$ Calling $c=\max_{x\in M}\abs{\tilde{v}_x}^2$, the choice \(\delta=\varepsilon/2c\) fulfills the desired inequality. Moreover,
\begin{equation}
t^\delta_{x_*}(v,v)=h_{x_*}(v,v)-\delta\,\tilde{v}_{x_*}^{\flat}(v)\tilde{v}_{x_*}^{\flat}(v)=-\delta\abs{v}^4<0,\qquad \forall\delta>0.
\end{equation}
Hence, every neighborhood of \(h\) contains a symmetric tensor that is not positive semidefinite. By the arbitrariness of \(h\), this implies
\begin{equation}
\mathrm{Deg}(M)\setminus\mathrm{Met}(M)\subset \partial\mathrm{Deg}(M).
\end{equation}

The converse is immediate: every metric \(g\in\mathrm{Met}(M)\) has a neighborhood completely contained in \(\mathrm{Met}(M)\), and so \(\mathrm{Met}(M)\) is the interior of \(\mathrm{Deg}(M)\). More precisely, we are going to prove that
\[\forall g\in\mathrm{Met}(M)\ \exists\varepsilon>0
\text{ such that }
\forall h\in\Gamma(\mathrm{S}^2T^*M)
\text{ with }d(g,h)<\varepsilon,\quad
h\in\mathrm{Met}(M).
\]
Under this hypothesis,
\begin{equation}
    \abs{h_x(v,v)-g_x(v,v)}\leq \abs{v}^2d(g,h)\leq \abs{v}^2\varepsilon, \qquad \forall x\in M,\quad v\in T_xM,
\end{equation}
and therefore
\[
g_x(v,v)-\varepsilon\abs{v}^2\leq h_x(v,v)\leq g_x(v,v)+\varepsilon\abs{v}^2.
\]
Hence, whenever
\begin{equation}
\varepsilon<\min_{x\in M}\min_{\substack{u\in T_xM\\ \abs{u}=1}}g_x(u,u),
\end{equation}
\(h\) is positive definite. Thus, providing the claim
\[\partial\mathrm{Deg}(M)=\mathrm{Deg}(M)\setminus\mathrm{Met}(M) \]
\end{proof}

\medskip

 We would like to claim the same in our setting: \(C^\infty(I,\mathrm{Deg}(\bS))\) is a convex cone in \(C^\infty(I,\Gamma(\mathrm{S}^2 T^*\bS))\) whose interior is \(C^\infty(I,\mathrm{Met}(\bS))\). However, this is not the case. Indeed, both norms \eqref{eq:norm-Sob} and \eqref{eq:norm-bdy} control the metrics via a Sobolev norm on the sphere and, with respect to this norm, \(\mathrm{Deg}(\bS)\) has no interior. The proof follows similar steps, but now the distance is induced by the Sobolev norm. Thus, we aim to prove that
\[\forall \varepsilon>0,\forall h\in \mathrm{Deg}(\bS),\,\exists t\notin \mathrm{Deg}(\bS) \text{ such that } d(h,t)< \varepsilon.\]
Namely, every open ball centered at a (possibly degenerate) metric contains a symmetric tensor that is not positive semidefinite.

Let us consider a point \(x_*\in\bS\) and a geodesic polar coordinate system \((r,\theta)\) centered at \(x_*\). The standard metric in these coordinates is
\begin{equation}
\hat{q}=dr^2+\sin^2r\,d\theta^2.
\end{equation}
Consider a one-parameter family of functions with compact support in the chart given by
\begin{equation}
u_{\delta}(r)=\begin{cases}
1 & r\leq \delta\\
\frac{\log(\frac{r_0}{r})}{\log(\frac{r_0}{\delta})} & \delta<r<r_0\\
0 & r\geq r_0
\end{cases}
\end{equation}
This family of functions tends to \(0\) in \(H^1(\bS)\) as \(\delta\to 0\). Indeed,
\begin{align*}
&\oint_{\bS}\abs{u_\delta}^2\epsilon_{\hat{q}}=2\pi \int_0^\delta\sin r\, dr+\frac{2\pi}{\abs{\log(\frac{r_0}{\delta})}^2}\int_\delta^{r_0}\abs{\log (\frac{r_0}{r})}^2\sin r\,dr\\
&\leq2\pi\delta+\frac{2\pi}{\log^2(\frac{r_0}{\delta})}\int_\delta^{r_0}\log^2 (\frac{r_0}{r})\,dr\\
&=2\pi\delta+\frac{2\pi}{\log^2(\frac{r_0}{\delta})}\left[r\left(\log^2(\frac{r_0}{r})+2\log(\frac{r_0}{r})+2\right)\right]_\delta^{r_0}\\
&=2\pi\delta+\frac{2\pi\delta}{\log^2(\frac{r_0}{\delta})}\left[2\frac{r_0}{\delta}-\log^2(\frac{r_0}{\delta})-2\log(\frac{r_0}{\delta})-2\right]
\end{align*}
and the logarithmic behavior also guarantees control over the derivatives:
\begin{align*}
&\oint_{\bS}\abs{\hat{\Dd}u_\delta}^2\epsilon_{\hat{q}}=2\pi \int_0^{r_0}\left(\frac{du_{\delta}}{dr}\right)^2\sin r\, dr\\
&=\frac{2\pi}{\log^2(\frac{r_0}{\delta})}\int_\delta^{r_0}\frac{\sin r}{r^2}dr\\
&\leq\frac{2\pi}{\log^2(\frac{r_0}{\delta})}\int_\delta^{r_0}\frac{d r}{r}\\
&=\frac{2\pi}{\log(\frac{r_0}{\delta})}
\end{align*}
Thus, given any differential one-form \(\alpha\in \Omega^1(\bS)\), we define \(t=h-u_{\delta}\alpha\otimes \alpha\). Up to a rescaling of \(\alpha\), we can always arrange that, for a given \(v\in T_{x_*}\bS\),
\[t_{x_*}(v,v)=h_{x_*}(v,v)-\alpha(v)^2<0,\]
and so \(t\notin \mathrm{Deg}(\bS)\). Moreover,
\begin{equation}
\begin{split}
&d(h,t)^2=\norm{u_\delta\alpha\otimes\alpha}^2_{H^1(\bS,\mathrm{S}^2T^*\bS)}\\
&=\oint_{\bS}\left(\abs{u_{\delta}\alpha\otimes\alpha}^2+\abs{\hat{\Dd}(u_\delta\alpha\otimes\alpha)}^2\right)\epsilon_{\hat{q}}\\
&\leq 2\oint_{\bS}\left(\abs{u_{\delta}\alpha\otimes\alpha}^2+u_\delta^2\abs{\hat{\Dd}(\alpha\otimes\alpha)}^2+\abs{\hat{\Dd}u_\delta}^2\abs{\alpha\otimes\alpha}^2\right)\epsilon_{\hat{q}}\\
&=2\oint_{\bS}u_\delta^2\left(\abs{\alpha}^4+2\abs{\hat{\Dd}\alpha}^2\abs{\alpha}^2\right)\epsilon_{\hat{q}}+2\oint_{\bS}\abs{\hat{\Dd}u_\delta}^2\abs{\alpha}^4\epsilon_{\hat{q}}\\
&\leq 2 \norm{u_{\delta}}_{H^1(\bS)}\max\left(\abs{\alpha}^4+2\abs{\hat{\Dd}\alpha}^2\abs{\alpha}^2\right).
\end{split}
\end{equation}
Hence, whenever \(\delta\) is sufficiently small, \(t\) is contained in a ball of radius \(\varepsilon\) in \(H^1(\bS,\mathrm{S}^2T^*\bS)\). This result automatically extends to our case. Let \(h\in C^\infty(I,\mathrm{Deg}(\bS))\) and consider
\begin{equation}
t(s)=h(s)-u_{\delta}\,\alpha\otimes \alpha
\end{equation}
where \(\alpha\) is chosen as above for some fixed \(s_*\in I\). Hence, \(t\notin C^\infty(I,\mathrm{Deg}(\bS))\). However, the term involving \(\alpha\) has no dependence on \(s\), so its derivatives vanish, and in both norms \eqref{eq:norm-Sob} and \eqref{eq:norm-bdy}, the difference reduce to
\begin{equation}
\norm{t-h}^2_{H^1}=\norm{u_\delta\alpha\otimes\alpha}^2_{H^1(\bS,\mathrm{S}^2T^*\bS)}(s_1-s_0)\,,\qquad \norm{t-h}^2=\norm{u_\delta\alpha\otimes\alpha}^2_{H^1(\bS,\mathrm{S}^2T^*\bS)}(s_1-s_0).
\end{equation}
Clearly, this behavior presents itself also in the case in which \(\Gamma(\mathrm{S}^2 T^*\bS)\) is equipped with the $L^2$ norm instead of the Sobolev one. Notice that $\mathrm{Deg}(\bS)$ is in fact closed in \(\Gamma(\mathrm{S}^2 T^*\bS)\) with respect to both the norms. By a common argument if $h_n\in \mathrm{Deg}(\bS)$ converges to $h\in\Gamma(\mathrm{S}^2 T^*\bS)$ in $L^2$ norm (and so in particular in Sobolev norm)  then there exists a subsequence $h_{n_k}$ converges pointwise to $h$ almost everywhere. Since all the functions are smooth by hypothesis then
\[\abs{h(x)-h_{n_k}(x)}^2\xrightarrow[k\to+\infty]{}0\,,\qquad \forall x\in\bS\]
Now, since $(h-h_{n_k})_x$ is a bilinear form on a finite-dimensional vector space $T_x\bS$ is bounded, hence there exists $c>0$ such that
\[|(h-h_{n_k})_x(v,v)|\leq c|(h-h_{n_k})_x|\abs{v}^2\qquad\forall v\in T_x\bS.\]
 Hence 
 \begin{equation}
     h_x(v,v)=\lim_{k\to+\infty}(h_{n_k})_x(v,v)\geq 0
 \end{equation}
providing
\[h\in \mathrm{Deg}(\bS)\] 

Clearly, this behavior also occurs when \(\Gamma(\mathrm{S}^2 T^*\bS)\) is equipped with the \(L^2\) norm instead of the Sobolev norm. Furthermore, notice that \(\mathrm{Deg}(\bS)\) is closed in \(\Gamma(\mathrm{S}^2 T^*\bS)\) with respect to both norms. Indeed, let \(h_n\in \mathrm{Deg}(\bS)\) be a sequence converging to \(h\in\Gamma(\mathrm{S}^2 T^*\bS)\) in the \(L^2\) norm. Then, by the standard subsequence argument, there exists a subsequence \(h_{n_k}\) that converges pointwise to \(h\) almost everywhere. In particular,
\[\abs{h(x)-h_{n_k}(x)}^2\xrightarrow[k\to+\infty]{}0\]
for almost every \(x\in\bS\). Since, for each \(x\), \((h-h_{n_k})_x\) is a bilinear form on the finite-dimensional vector space \(T_x\bS\), there exists a constant \(c>0\) such that
\[\abs{(h-h_{n_k})_x(v,v)}\leq c\abs{(h-h_{n_k})_x}\abs{v}^2\qquad\forall v\in T_x\bS.\]
Therefore, for almost every \(x\in\bS\),
\begin{equation}
    h_x(v,v)=\lim_{k\to+\infty}(h_{n_k})_x(v,v)\geq0\qquad\forall v\in T_x\bS.
\end{equation}
Hence \(h\) is positive semidefinite almost everywhere. Since \(h\) is smooth by hypothesis, the map
\[x\mapsto h_x(v_x,v_x)\]
is continuous for every smooth vector field \(v\), and therefore positivity almost everywhere implies positivity everywhere. Thus
\[h\in\mathrm{Deg}(\bS).\]
Consequently, \(\mathrm{Deg}(\bS)\) is closed with respect to the \(L^2\) norm. Since convergence in the Sobolev \(H^1\) norm implies convergence in the \(L^2\) norm, it is also closed with respect to the Sobolev norm.

\bibliographystyle{JHEPs}
\bibliography{biblio}

\end{document}